\documentclass[fleqn,usenatbib,useAMS]{mnras}

\usepackage{graphicx}	
\usepackage{amsmath}	
\usepackage{amssymb}	
\usepackage{multicol}   
\usepackage{bm}		    
\usepackage{pdflscape}	
\usepackage{lscape}     
\usepackage{float}      

\newcommand{\pcmc}{\,cm$^{-3}$}	
\newcommand{\kms}{\,km\,s$^{-1}$} 
\newcommand{\python}{{\sc python~}}
\newcommand{\pythonstop}{{\sc python}}
\newcommand{\Ha}{H$\alpha$~}
\newcommand{\Hastop}{H$\alpha$}
\newcommand{\Hb}{H$\beta$~}
\newcommand{\Hbstop}{H$\beta$}
\newcommand{\degrees}{$^{\circ}$}
\newcommand{\atomictransition}[2]{#1 \textsc{#2}}

\defcitealias{Parkinson2020}{P20}

\usepackage[T1]{fontenc}
\usepackage{ae,aecompl}
\usepackage{todonotes}

\usepackage{newtxtext,newtxmath}

\title[Reprocessing in tidal disruption events]{Optical line spectra of tidal disruption events from reprocessing in optically thick outflows}

\author[E. J. Parkinson et al.]
{Edward J. Parkinson$^{1}$\thanks{E-mail: \href{mailto:ejp1n17@soton.ac.uk}{e.j.parkinson@soton.ac.uk}}, 
Christian Knigge$^{1}$,
James H. Matthews$^{2}$,
Knox S. Long$^{3,~4}$,\newauthor
Nick Higginbottom$^{1}$,
Stuart A. Sim$^{5}$
and Samuel W. Mangham$^{1}$
\\
$^{1}$ Department of Physics and Astronomy, University of Southampton, Southampton, SO17 1BJ, UK\\
$^{2}$Institute of Astronomy, University of Cambridge, Madingley Road, Cambridge CB3 0HA, UK\\
$^{3}$Eureka Scientific Inc., 2542 Delmar Avenue, Suite 100, Oakland, CA, 94602-3017, USA\\
$^{4}$Space Telescope Science Institute, 3700 San Martin Drive, Baltimore, MD, 21218, USA\\
$^{5}$School of Mathematics and Physics, Queen's University Belfast, University Road, Belfast, BT7 1NN, UK\\
}

\date{Accepted XXX. Received YYY; in original form ZZZ}
\pubyear{2020}

\begin{document}
\label{firstpage}
\pagerange{\pageref{firstpage}--\pageref{lastpage}}
\maketitle


\begin{abstract}
A significant number of tidal disruption events (TDEs) radiate primarily at optical and ultraviolet (UV) wavelengths, with only weak soft X-ray components. One model for this optical excess proposes that thermal X-ray emission from a compact accretion disc is reprocessed to longer wavelengths by an optically thick envelope. Here, we explore this reprocessing scenario in the context of an optically thick accretion disc wind. Using state-of-the-art Monte Carlo radiative transfer and ionization software, we produce synthetic UV and optical spectra for wind and disc-hosting TDEs. Our models are inspired by observations, spanning a realistic range of accretion rates and wind kinematics. We find that such outflows can efficiently reprocess the disc emission and produce the broad Balmer and helium recombination features commonly seen in TDEs and exhibit asymmetric red wings. Moreover, the characteristic colour temperature of the reprocessed spectral energy distribution (SED) is much lower than that of the accretion disc. We show explicitly how changes in black hole mass, accretion rate and wind properties affect the observed broadband SED and line spectrum. In general, slower, denser winds tend to reprocess more radiation and produce stronger Balmer emission. Most of the outflows we consider are too highly ionized to produce UV absorption features, but this is sensitive to the input SED. For example, truncating the inner disc at just $\simeq 4 ~ R_{\text{ISCO}}$ lowers the wind ionization state sufficiently to produce UV absorption features for sight lines looking into the wind.
\end{abstract}

\begin{keywords}
accretion, accretion discs -- black hole physics -- galaxies: nuclei -- transients: tidal disruption events
\end{keywords}


\section{Introduction} \label{sec: introduction}

    Tidal disruption events (TDEs) occur when a star passes within the tidal radius $R_{t} = R_{*} (\text{M}_{\text{BH}}/\text{M}_{*})^{1/3}$ of a supermassive blackhole (SMBH), where $R_{*}$ and $\text{M}_{*}$ are the radius and mass of the star, respectively, and $\text{M}_{\text{BH}}$ is the mass of the SMBH. Within this radius, the star's self gravity is overwhelmed by tidal forces, causing the star to be torn apart \citep{Hills1975}. Roughly half of the stellar debris is accreted onto the SMBH and powers a luminous flare, whilst the rest of the material is ejected from the system \citep{Rees1988, Phinney1989}. 
    
    The physical processes involved in generating the luminous flare are still uncertain. In the simplest, "canonical", scenario, the flare is a consequence of bound stellar material forming a quasi-circular accretion disc through which material is transported at typically super-Eddington rates \citep[see, e.g.][]{Shiokawa_2015, Hayasaki2016}. The light observed is then thermal emission from this hot accretion disc. The emission from TDEs is expected to be dominated by soft X-ray emission, with a weak optical and ultraviolet (UV) component \citep{Rees1988, Strubbe2009}. 
    
    \begin{figure*}
        \centering
        \includegraphics[scale=0.58]{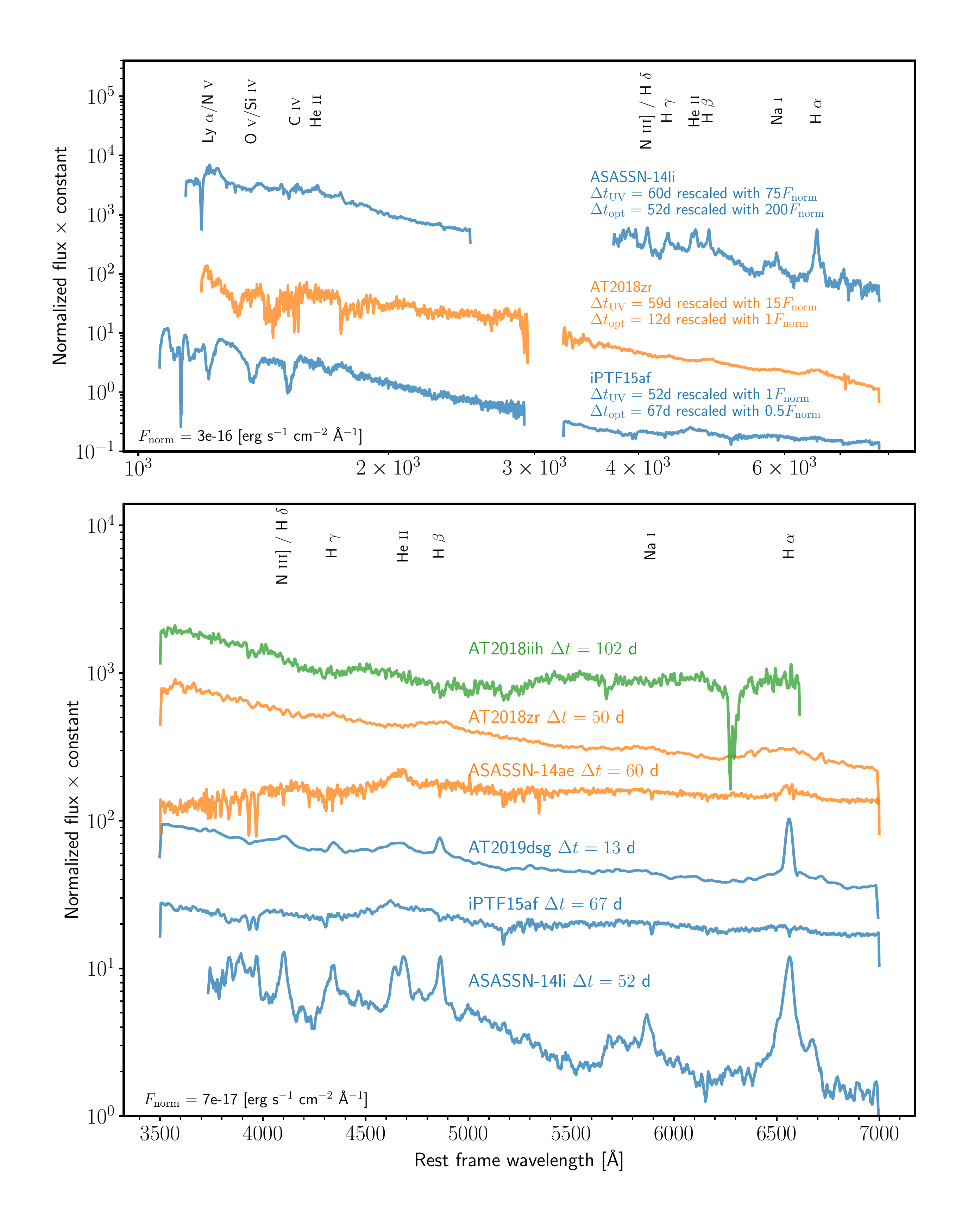}
        \caption{Rest frame optical and UV spectra for a sample of TDEs. The spectra have been coloured depending on their \citet{VanVelzen2020} spectral class. Blue corresponds to TDE-Bowen, orange to TDE-H and green to TDE-He. \textit{Top}: normalized UV and optical spectra of ASASSN-14li \citep{Holoien2016}, AT2018zr \citep{Hung2019} and iPTF15af \citep{Blagorodnova_2019}. The optical spectra have been re-scaled for the blue edge to roughly match the red edge of the UV spectra. The re-scaling amounts, as well as observation phase of the spectra are labelled in the figure. Specifically visible is how some TDEs display BELs (e.g. ASASSN-14li), whilst others exhibit BALs (e.g. iPTF15af). \textit{Bottom}: optical spectra of AT2018iih \citep{VanVelzen2020}, AT2018zr \citep{Hung2019}, ASASSN-14ae \citep{Holoien2014}, ASASSN-14li \citep{Holoien2016}, AT2019dsg \citep{VanVelzen2020}, iPTF15af \citep{Blagorodnova_2019} and ASASSN-14li \citep{Holoien2014}. Important transitions have been labelled at the top of the figure. None of the spectra have been host galaxy subtracted and have been smoothed using a boxcar filter.}
        \label{fig: dual_p_tde_uv_opt_obs}
    \end{figure*}
    
    Contrary to this expectation, an increasing number of UV and optically bright TDEs with a weaker than expected soft X-ray component have been discovered over the past decade. The exact source of the optical emission is still uncertain, but it is clearly inconsistent with the spectral energy distribution (SED) of a hot accretion disc. Two mechanisms have been proposed to explain this ``optical excess''. In the first, the optical emission is powered by shocks associated with collisions between infalling stellar debris streams \citep[e.g.][]{Dai_2015, Piran2015, Shiokawa_2015}. {In the second, the optical excess is due to the reprocessing -- and consequent softening -- of the disc's X-ray radiation in a surrounding optically thick medium, such as an outflow \citep[e.g.][]{Strubbe2009, Metzger2016a, Roth2016, Dai_2008, roth_what_2018, Lu_2020, Piro_2020, bonnerot_2020}.}
    
    Observationally, the optical spectra of TDEs are characterised by broad (FWHM $\sim 10^{4} \text{ km s}^{-1}$) hydrogen and/or helium lines (see bottom panel of Figure \ref{fig: dual_p_tde_uv_opt_obs}). In some TDEs, a complex series of emission lines due to Bowen fluorescence are also present. In the taxonomy proposed by \citet {VanVelzen2020}, TDEs can be broadly split into three distinct sub classes: i) TDE-H: broad \Ha and \Hb emission lines; ii) TDE-Bowen: broad \Ha and \Hb emission lines with a complex of emission lines around \atomictransition{He}{ii} $\lambda 4686$ with most due to Bowen fluorescence \citep[see][]{Bowen1934, Bowen1935}, and; iii) TDE-He: no broad Balmer features but a broad emission feature near \atomictransition{He}{ii} $\lambda 4686$. The most common class of optical TDE (to date) is TDE-H, followed by TDE-Bowen and TDE-He.
    
    The discovery of blue-shifted broad absorption lines (BALs) in the ultraviolet spectra of some TDEs \citep[e.g.][]{Cenko_2016, Blagorodnova_2019, Hung2019, hung_discovery_2020} provides unambiguous observational evidence for the presence of fast and powerful outflows in these systems (top panel of Fig. \ref{fig: dual_p_tde_uv_opt_obs}). Additional evidence for powerful sub-relativistic outflows in TDEs also comes from blueshifted broad \textit{emission} lines (BELs) \citep{arcavi14, roth_what_2018, Hung2019, Nicholl2020} and from radio observations \citep{vanvelzen16, alexander16, alexander17, anderson20}. Not all TDEs exhibit BALs at the  same stage in their outburst evolution, however. For example, ASASSN-14li instead exhibits BELs (cf. the top panel of Fig. \ref{fig: dual_p_tde_uv_opt_obs}). Acknowledged by, e.g., \citet{Blagorodnova_2019, hung_discovery_2020}, the BEL vs. BAL dichotomy of TDEs is reminiscent of Type I quasars (QSOs). Most QSOs exhibit BELs in their UV spectra, but $\simeq 20$ per cent of the population instead display prominent BALs \citep[the so-named Broad Asborption Line QSOs][]{Weymann1991, Dai_2008, Knigge2008, Allen2010}. In QSOs, line formation within outflows has, in the past, been invoked to explain the BEL vs. BAL dichotomy, following as a consequence of the orientation of an observer \citep[e.g.][]{Murray1995, Elvis2000, Higginbottom2013}.
    
    In \citet[][]{Parkinson2020}, hereafter referred to as \citetalias{Parkinson2020}, we showed that line formation in an accretion disc wind could explain the existence of both BELs and BALs in the UV spectra of TDEs. Specifically, we proposed that the BEL vs. BAL dichotomy arises naturally as a consequence of the orientation of an observer. We found that BALs are preferentially seen for sight lines looking into the outflow, whereas BELs are more likely to be observed for viewing angles which are above or below the wind cone. The work presented here extends our modelling to longer wavelengths. We find that accretion disc emission reprocessed by an accretion disc wind is a promising mechanism for the formation of the distinctive \textit{optical} emission line spectra seen in TDEs. Our goal, then, is to test if the outflow reprocessing model, for the optical excess in TDEs, can naturally produce the optical (and ultraviolet) emission line spectra and broadband SED seen in TDEs. The specific type of outflow we consider is a simple biconical accretion disc wind. This is quite a natural scenario, since strong, radiation-driven mass loss is almost inevitable, given the extreme luminosities of TDEs.
    
    The paper is structured as follows. In Section~\ref{sec: method}, we describe our wind models and radiative transfer calculations. We then present the results of our modelling in Section~\ref{sec: results} and discuss their implications in Section~\ref{sec: discussion}. Finally, we summarize our findings in Section~\ref{sec: conclusion}.

\section{Radiative transfer and model setup} \label{sec: method}

    \subsection{Radiative transfer and ionization}
        Numerical simulations were conducted using \pythonstop\footnote{\python is a collaborative open-source project available at \hyperlink{https://github.com/agnwinds/python}{github.com/agnwinds/python}.}, a state-of-the-art Monte Carlo radiative transfer and ionization code for moving media using the Sobolev approximation \citep[e.g.][]{Sobolev1957, Rybicki1978}. \python was originally described by \citet{Long2002} and subsequent improvements to the code have been described multiple times in the literature \citep{Higginbottom2013, Higginbottom2014, Matthews2015, Matthews2016a}. Here, we provide only a brief description.
        
        \subsubsection{Basic operation}
        \python consists of two separate calculation stages. The first stage concerns itself with calculating the ionization state, level populations and temperature structure of an outflow on a spatially discretised grid. This is done iteratively by tracking a population of Monte Carlo energy quanta (``photon packets'') and simulating their random walk through the grid. Photon packets are randomly generated over a wide frequency range, sampled from the spectral energy distribution (SED) of the radiation sources included in the simulation. As photon packets travel through the grid, they interact with the plasma and update Monte Carlo estimators which are used to describe the radiation field in each grid cell. The heating effect of photon packets is recorded and used to iterate the temperature towards thermal equilibrium, where the amount of heating and cooling in each grid cell is eventually balanced.
        
        Once the photon packets have been transported through the grid, updated temperature and radiation field estimators are used to recalculate level populations and an updated ionization state of the outflow. This process is repeated until the simulation has converged. We consider a grid cell to be converged when i) the electron and radiation temperature have stopped changing between iterations to within 5 per cent, and, ii) when the heating and cooling rates are balanced to within 5 per cent. It is usually not necessary, or expected, for all grid cells to converge. Typically, cells with poor photon statistics or noisy Monte Carlo estimators tend to not converge. These cells are usually located near the outer edge of the computational domain and are relatively unimportant; they typically contribute little to the final result.
        
        The second calculation stage produces synthetic spectra for a converged simulation. Additional populations of photon packets are generated, typically over a narrow frequency range to give high signal-to-noise, and are flown through the \textit{converged} grid to generate spectra for a selection of sight lines.
        
        \subsubsection{Atomic data}
        By default \python assumes solar abundances in the modelled outflow, following \cite{1994A&AS..108..287V}. We use the same atomic data as outlined by \citet{Long2002}, with subsequent improvements as described by \citet{Higginbottom2013} and \citet{Matthews2015}. Hydrogen and helium are described with a multi-level model atom, treated using the ``macro-atom'' formalism of \citet{Lucy2002, Lucy2003}. Metals, however, are treated with the two-level atom formalism described by \citet{Long2002}. The resulting hybrid approach is explained by \citet{sim_two-dimensional_2005} and \citet{Matthews2015}.
        
        \subsubsection{Radiation sources} \label{sec: radiation_sources}
        Multiple radiation sources can be included in a \python simulation. In the models presented here, we include two radiation sources: an accretion disc and the outflow itself. However, the accretion disc is the only \textit{net} source of photons, as all emission produced in the outflow is reprocessed disc radiation. 
        
        Modulo adiabatic cooling, the outflow is assumed to be in radiative equilibrium. Any energy absorbed by it is reprocessed and either re-radiated or converted to $PdV$ work. {Radiative reprocessing in the outflow takes place via radiative recombination, free-free and line emission and Compton scattering (which is also included in the heating/cooling balance of the plasma).} As these processes depend on the temperature and ionization state, the number of photon packets generated by the outflow is updated as these quantities change. The accretion disc is assumed to be geometrically thin and optically thick. It it treated as an ensemble of annuli, each radiating as a black body with a standard $\alpha$-disc temperature profile \citep{Shakura1973}. The disc SED is therefore specified entirely by the mass accretion rate, $\dot{\text{M}}_{\text{disc}}$, and the mass of the black hole, $\text{M}_{\text{BH}}$. In principle, both the inner and outer radii of the disc are free parameters, but the former is typically set to the innermost stable circular orbit for a Schwarzschild black hole. Both fore-shortening and limb-darkening are included, resulting in a highly anisotropic radiation field. 
        
        In reality, the inner disc region in TDEs are likely radiation dominated and vertically extended. However, the structure, evolution and stability of such discs is an area of intense research and still highly uncertain \citep[e.g.][]{Hirose2009, Jiang_2013, blaes14, Shen_2014}. In the near-Eddington critical accretion regime, where $\dot{\text{M}}/\dot{\text{M}}_{\text{Edd}} \approx 1$, so-called ``slim'' disc models are often used to describe the disc structure \citep[see, e.g.][]{Abramowicz1988, Strubbe2009}. In these slim disc models, for large Eddington fractions the temperature for the inner disc region is lowered due to decreasing radiative efficiency and consequently increased radiation trapping. In the accretion regime of interest, i.e. $\dot{\text{M}}/\dot{\text{M}}_{\text{Edd}} \leq 0.5$, we find that both a slim disc and a standard $\alpha$-disc temperature profile result in very similar disc SEDs. Our reliance on a simple $\alpha$-disc treatment, then, is one borne out of practicality in the absence of a more physically realistic description.
        
        \subsubsection{Clumping}
        Whilst \python was originally developed to model smooth outflows, in reality various instabilities within the flow are likely to break the smooth flow up into clumps. Multiple clumping mechanisms have already been identified in various astrophysical settings \citep[e.g.][]{Owocki1983, McCourt2018, dannen_clumpy_2020}. From a computational standpoint, addressing this is difficult. Not only does clumping introduce additional parameters into already complex models, but, ideally, the spatial resolution of the computational grid would be high enough to resolve each individual clump. This is not yet feasible, and so \python implements a necessarily simple approximation known as \textit{micro-clumping}, which is commonly used in stellar-wind modelling \citep[e.g.][]{Hamann1998, Hamann2013}. Micro-clumping in \python has been discussed in previous work by \citet{Matthews2016a, Matthews2020} and, in the context of the problem at hand by \citetalias{Parkinson2020}. Within the micro-clumping framework, clumps are assumed to be optically thin, smaller than all relevant length-scales and embedded in a vacuum. In this limit, clumping can be parameterised by a single parameter $f_{v}$, the volume filling factor. The density of a clump is enhanced by a factor $D = 1 / f_{v}$ relative to the density of the equivalent smooth flow.  Consequently, at a fixed temperature, processes which scale linearly with density (e.g. electron scattering) remain unchanged, whereas processes which scale with the square of the density (e.g. recombination) are enhanced.
        
        One of the long-standing challenges for wind models in QSOs is overionization. This can occur when an outflow is exposed to an (E)UV or X-ray radiation field near the central engine \citep{Proga2002, Higginbottom2013}. In line-driven winds, if the wind is too highly ionized it can become impossible to sustain the wind \citep[e.g.][]{Proga2000, Proga2002}. Irrespective of the driving mechanism of the wind, overionization can also prevent the formation of both absorption and emission features. Clumping is one natural way to overcome the so-called ``overionization problem'' and can moderate the ionization state of the outflow (e.g. \citealt{Hamann2013, Matthews2016a}; another solution is self-shielding, e.g. \citealt{Murray1995,Proga2000, Proga2004}). Clumping is therefore also an important effect to consider in TDEs. {In previous QSO models, \citet{Matthews2016a} and \citet{Matthews2020} assumed clumping factors of both $f_{v} = 0.1$ and $f_{v} = 0.01$.} Throughout this work we have adopted a volume filling factor $f_{v} = 0.1$, corresponding to a density enhancement by a factor of ten. 

        \subsubsection{Special relativity}
        Prior to this work, \python made the assumption that the outflow velocities were small compared to the speed of light. Doppler shifts were therefore treated "classically" (proportional to linear order in $v/c$), and no other corrections associated with relativistic effects were considered. For this work, however, we have updated \python to fully incorporate special relativistic effects. \python now explicitly distinguishes and correctly transforms between the co-moving (local/fluid) and observer frames. We ensure that energy is conserved in the co-moving frame and that the correct special relativistic transforms are applied when converting between frames \citep[see, e.g.,][]{Castor1972, Mihalas_Mihalas1984}. 
        
        The synthetic spectra are generated in the observer frame, the same frame in which photons are transported. Photon interactions, however, are handled in the co-moving frame. This means photons are transformed into the local fluid frame when they interact with the outflow, e.g. during resonance scattering. All atomic processes, as well as heating and cooling, are treated in the co-moving frame.

    \subsection{A biconical wind model}
    
        \begin{figure}
            \centering
            \includegraphics[scale=0.35]{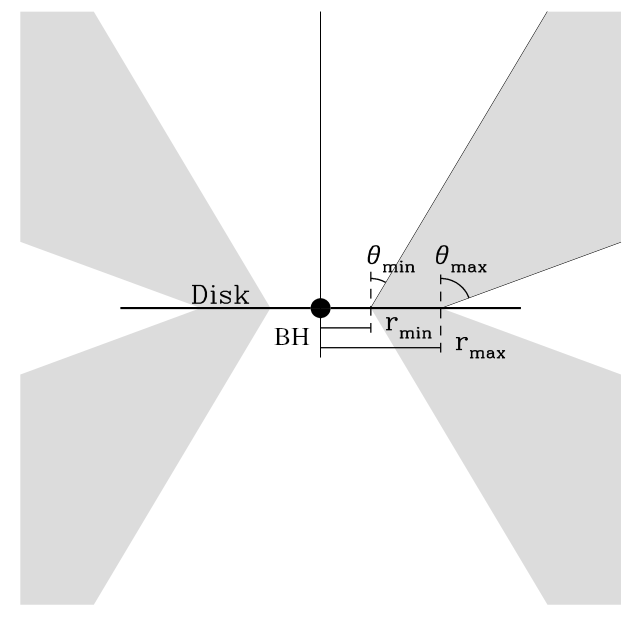}
            \caption{A cartoon showing the basic geometry of the \citet{Shlosman1993} biconical disc wind model, adapted from \citet{Long2002}.}
            \label{fig: sv93_cartoon}
        \end{figure}
        
        In this work, we model a disc wind outflow using a kinematic biconical wind prescription presented originally by \citet{Shlosman1993}. This model is illustrated in Figure \ref{fig: sv93_cartoon}. In this prescription, streamlines emerge from disc radii between $r_{\text{min}}$ and $r_{\text{max}}$ at angles relative to the disc normal given by
        \begin{equation}
            \theta(r_{0}) = \theta_{\text{min}} + (\theta_{\text{max}} - \theta_{\text{min}}) x^{\gamma}, 
            \label{eq: sv93_theta}
        \end{equation}
        where
        \begin{equation}
            x = \frac{r_{0} - r_{\text{min}}}{r_{\text{max}} - r_{\text{min}}}.
            \label{eq: sv93_x}
        \end{equation}
        Here, $r_{0}$ is the launch radius of a streamline, $\theta_{\text{min}}$ and $\theta_{\text{max}}$ are the minimum and maximum opening angles of the wind, and $\gamma$ controls the concentration of streamlines towards either of the two boundaries $r_{\text{min}}$ and $r_{\text{max}}$.
        
        The launch velocity, $v_{0}$, and terminal velocity, $v_{\infty}$, of a streamline are set to the local sound
        speed $c_{s}$ and a multiple of the escape velocity, $v_{\rm esc}$, at the streamline footpoint, respectively. The poloidal velocity, $v_{l}$, at a distance $l$ along a streamline, is given by the following power law,
        \begin{equation}
            v_{l}(r_{0}) =  v_{0} + (v_{\infty} - v_{0})\left[ \frac{(l / R_{v})^{\alpha}}{(l / R_{v})^{\alpha} + 1} \right],
            \label{eq: sv93_velocity}
        \end{equation}
        where $R_{v}$ is the acceleration length scale, and $\alpha$ controls the acceleration along the streamline. The rotational velocity, $v_{\phi}$, is Keplerian at the footpoint of a streamline and is assumed to conserve specific angular momentum as it rises above the disc. At a cylindrical radius $r$, $v_{\phi}$ is given by
        \begin{equation}
            v_{\phi}(r) = v_{k} r_{0}/r,
            \label{eq: sv93_rotational_vel}
        \end{equation}
        where $v_{k}$ is the Keplerian velocity at $r_{0}$. The density at any point in the wind is obtained by enforcing mass continuity, giving 
        \begin{equation}
            \rho(r, z) = \frac{r_{0}}{r} \frac{dr_{0}}{dr} \frac{\dot{m}^{\prime}}{v_{z}(r, z)},
        \end{equation}
        where $\dot{m}^{\prime}$ is the mass-loss rate per unit surface area of the accretion disc, 
        \begin{equation}
            \dot{m}^{\prime} \propto \dot{\text{M}}_{\text{wind}} r_{0}^{\lambda} \cos[\theta(r_{0})]. 
            \label{eq: sv93_mass_loss_disc_surf}
        \end{equation}
        The parameter $\lambda$ controls where mass is lost from the disc, and $\dot{\text{M}}_{\text{wind}}$ is the total mass loss rate of the wind. A value of $\lambda = -2$ results in roughly uniform wind densities at the base of the outflow, across the disc surface. 
    
    \subsection{Simulation grid setup} \label{sec: model_grid_setup}

        \begin{table}
            \centering
            \resizebox{\columnwidth}{!}{		\begin{tabular}{ccccc}
			\hline
			\multicolumn{5}{l}{Grid parameters}                                                                                                                    \\
			\hline
			Parameters                               & \multicolumn{3}{c}{Values} & Units                                                                          \\
			\hline
			$\dot{\text{M}}_{\text{disc}}$           & 0.05                       & 0.15                  & 0.50                  & $\dot{\text{M}}_{\text{Edd}}$  \\
			$\dot{\text{M}}_{\text{wind}}$           & 0.1                        & 0.3                   & 1.0                   & $\dot{\text{M}}_{\text{disc}}$ \\
			$v_{\infty}$                             & 0.1                        & 0.3                   & 1.0                   & $v_{\text{esc}}$               \\
			\hline
			\multicolumn{5}{l}{Geometry parameters}                                                                                                                \\
			\hline
			Parameters                               & \multicolumn{3}{c}{Values} & Units                                                                          \\
			\hline
			$\text{M}_{\text{BH}}$                   & $10^{6}$                   & $3 \times 10^{6}$     & $10^{7}$              & $\text{M}_{\odot}$             \\
			\hline
			$\text{R}_{\text{co}}, r_{\text{min}}$   & $8.85 \times 10^{11}$      & $2.66 \times 10^{12}$ & $8.85 \times 10^{12}$ & cm                             \\
													 & 6                          & 6                     & 6                     &
			$\text{R}_{\text{g}}$                                                                                                                                  \\
			$\text{R}_{\text{disc}}, r_{\text{max}}$ & $7 \times 10^{13}$         & $10^{14}$             & $1.5 \times 10^{15}$  & cm                             \\
													 & 473                        & 228                   & 102                   &
			$\text{R}_{\text{g}}$                                                                                                                                  \\
			$\alpha$                                 & 4                          & 4                     & 4                     & -                              \\
			$R_{v}$                                  & $3\times10^{14}$           & $8.9 \times 10^{14}$  & $3\times10^{15}$      & cm                             \\
													 & $2000$                     & $2000$                & $2000$                &
			$\text{R}_{\text{g}}$                                                                                                                                  \\
			$\theta_{\text{min}}$                    & 20                         & 20                    & 20                    & \degrees                       \\
			$\theta_{\text{max}}$                    & 65                         & 65                    & 65                    & \degrees                       \\
			$\text{R}_{\text{max}}$                  & $5 \times 10^{17}$         & $5 \times 10^{17}$    & $5 \times 10^{17}$    & cm                             \\
													 & $3.39 \times 10^{6}$       & $1.13 \times 10^{6}$  & $3.39 \times 10^{5}$  &
			$\text{R}_{\text{g}}$                                                                                                                                  \\
			$\gamma$                                 & 1                          & 1                     & 1                     & -                              \\
			$\lambda$                                & 2                          & 2                     & 2                     & -                              \\
			$f_{v}$                                  & 0.1                        & 0.1                   & 0.1                   & -                              \\
			\hline
			\multicolumn{5}{l}{Eddington ratio conversions}                                                                                                        \\
			\hline
			Ratio                                    & \multicolumn{3}{c}{Values} & Units                                                                          \\
			\hline
			$\text{M}_{\text{BH}}$                   & $10^{6}$                   & $3 \times 10^{6}$     & $10^{7}$              & $\text{M}_{\odot}$
			\\
			\hline
			0.05 $\dot{\text{M}}_{\text{Edd}}$       & $1.11 \times 10^{-3}$      & $3.33 \times 10^{-3}$ & $1.11 \times 10^{-2}$ &
			$\text{M}_{\odot}~\text{yr}^{-1}$                                                                                                                      \\
			0.15 $\dot{\text{M}}_{\text{Edd}}$       & $3.33 \times 10^{-3}$      & $9.99 \times 10^{-3}$ & $3.33 \times 10^{-2}$ &
			$\text{M}_{\odot}~\text{yr}^{-1}$                                                                                                                      \\
			0.50 $\dot{\text{M}}_{\text{Edd}}$       & $1.11 \times 10^{-2}$      & $3.33 \times 10^{-2}$ & $1.11 \times 10^{-1}$ &
			$\text{M}_{\odot}~\text{yr}^{-1}$                                                                                                                      \\
			\hline
		\end{tabular}}
            \caption{{Key parameters for the models presented. The top sub-table shows the values for the parameter grid as described in Section \ref{sec: model_grid_setup}, whilst the middle sub-table  defines the parameters which control the geometry for each black hole mass. The bottom table shows conversion values of the Eddington accretion rate into solar masses per year for each black hole mass.}}
            \label{table: key_parameters}
        \end{table}
        
        In our previous paper we explored the formation of UV lines in accretion disc winds associated with TDEs \citepalias{Parkinson2020}. In this paper, we now present both optical and UV spectra from a broader parameter search. 
        
        The masses of the black holes in TDEs are expected to fall within a broad, but well defined mass range. If the black hole is too large, i.e. $\gtrsim 10^{8}~\text{M}_{\odot}$, then the disruption radius will be within the event horizon of the black hole. In this case the star will be completely swallowed before it is disrupted \citep{MacLeod_2012}, resulting in no observable TDE. At the opposite end of the mass scale, the lowest inferred black hole masses are $\simeq 5 \times 10^{5}~\text{M}_{\odot}$ \citep{Wevers_2017, Mockler_2019, ryu2020, Zhou_2021}. It is worth noting that several different methods are being used to estimate black hole masses in TDEs. For example, if the velocity dispersion can be measured from the stellar kinematics of the host galaxy, the $\text{M}_{\text{BH}}-\sigma$ relation can be used \citep[e.g.][]{Wevers_2017}. Alternatively, a multi-parameter model can be fit to the observed light curve to extract the physical properties of the TDE \citep{Guillochon_2018, Mockler_2019}.\footnote{The widely used {\sc mosfit} light curve modelling package is available at \hyperlink{https://github.com/guillochon/MOSFiT}{https://github.com/guillochon/MOSFiT}}
        
        In order to settle on a reasonable range of black hole masses and Eddington ratios for our simulation grid, we used the largest compilation of these parameters by \citep{ryu2020}. Their estimates are based on {\sc tdemass}\footnote{\hyperlink{https://github.com/taehoryu/TDEmass}{https://github.com/taehoryu/TDEmass}} \citep{ryu2020}, which assumes a \textit{slow circularization} scenario for the in-falling stellar debris. In this model, the optical emission is generated from stream-stream collisions; both black hole and stellar mass can then be estimated from the optical luminosity and colour temperature at peak flare. Even though their underlying physical picture is quite different from that envisaged in our simulations -- where the reprocessed accretion disc radiation dominates the optical light -- the black hole masses inferred via this method are in agreement with those obtained by \textsc{mosfit}, in which the luminosity is assumed to be due to a disc \citep{Mockler_2019}.
        
\begin{figure}
            \centering
            \includegraphics[scale=0.52]{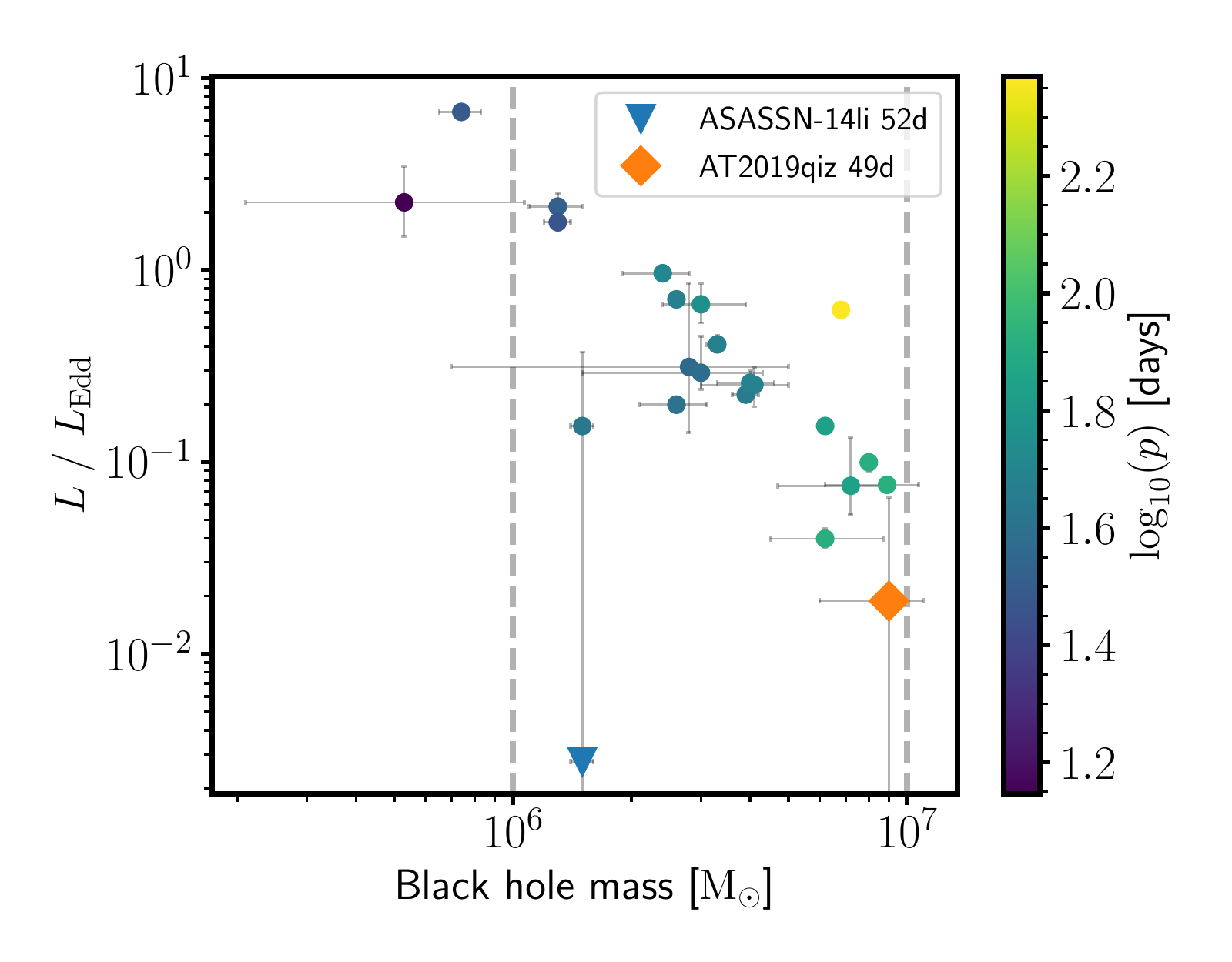}
            \caption{The Eddington luminosity fraction against inferred black hole masses for a sample of 20 TDEs taken from \citet{ryu2020} and the spectra included in Fig. \ref{fig: tde_model_comparison}, labelled in the legend. Each point from the \citet{ryu2020} sample has been coloured depending on the mass return time $p$ since disruption, measured in days. The Eddington luminosity is estimated using the inferred black hole masses. The bounded region represents the black hole mass parameter space covered by the grid in this work.}
            \label{fig: Mbh_parameter_space}
        \end{figure}

        Fig. \ref{fig: Mbh_parameter_space} shows the black hole masses and Eddington fractions for a sample of 20 TDEs, as provided by \citet{ryu2020}. We also include the TDE spectra shown in Figure \ref{fig: tde_model_comparison}, at the times the spectra were obtained. Based on Fig. \ref{fig: Mbh_parameter_space}, we consider three black hole masses in our simulations: $\text{M}_{\text{BH}} = 10^{6},~3\times10^{6},\text{ and } 10^{7}~\text{M}_{\odot}$. These cover the vast majority of the observationally inferred black hole parameter space.
        
        {
        In terms of accretion rates, we only consider the sub-Eddington regime in the present work: the highest accretion we consider is $\dot{\text{M}}_{\text{acc}} = 0.5 ~ \dot{\text{M}}_{\text{Edd}}$. Early on in their evolution, some TDEs with $\text{M}_{\text{BH}} \lesssim 10^{7}~\text{M}_{\odot}$ are expected to undergo a super-Eddington accretion phase, transitioning eventually to sub-critical accretion rates as the fallback rate of stellar debris diminishes {\citep{Strubbe2009, Wu2018, roth_radiative_2020}}. However, most TDEs in the \citet{ryu2020} sample appear to be sub-critical even at peak. Moreover, the TDE spectra to which we compare our models have typically been obtained $\simeq 2$ months post-peak (cf Fig.~\ref{fig: dual_p_tde_uv_opt_obs}). Many super-Eddington TDEs are expected to have transitioned to the sub-critical regime by this point \citep{Strubbe2009}. Last, but not least, the structure and radiative properties of super-Eddington accretion discs are still poorly understood. By restricting ourselves to the sub-Eddington regime we are able to avoid some of the resulting uncertainties.} We have made the assumption that all the luminosity generated in the TDE is (initially) emitted by the accretion disc. For the model with the largest accretion rate (with $\text{M}_{\text{BH}} = 10^{7}~\text{M}_{\odot}$), this places the maximum disc luminosity of the grid at $L_{\text{disc}} \approx 5 \times 10^{44}~\text{ergs s}^{-1}$.

        We not only explore the black hole mass dependence on the optical spectra with our simulation grid, but also the effects of three key parameters: (1) the disc accretion rate, (2) the wind mass-loss rate, and, (3) the terminal velocity of wind streamlines. For each parameter, we explore three values for three black hole masses. The values of these parameters are contained within the upper section of Table \ref{table: key_parameters}.
        
        The inner radius of the accretion disc in each model is set to the location of the innermost stable circular orbit (ISCO). {The outer disc radius is set to $R_{\text{disc}} = 10 R_{t}$, inspired by hydrodynamic simulations that follow the disruption of a solar-like star \citep{bonnerot_2020}.} The outflow is wide-angled and characterised by opening angles $\theta_{\text{min}} = 20^{\circ}$ and $\theta_{\text{max}} = 65^{\circ}$, emanating from the entire accretion disc surface. {The mass-loss rate per unit area from the disc is slightly enhanced at larger disc radii ($\lambda = 2$), meaning the density at the base of the wind increases with radius.}
        
        {The acceleration length scale for each wind model is set to $2000~R_{g}$, a value motivated by the radiation-hydrodynamic simulations presented by \citet{Dai2018}. In their model, the outflow is still accelerating at distances of $\sim 1500-3000~R_{g}$. The launch velocities of the streamlines are set to the the local sound speed at the footpoint of the streamline and are uniformly spaced at the base of the outflow ($\gamma = 1$).}
        
        {The outflow is logarithmically discretised onto a 2D cylindrical rotationally symmetric grid. Photons are transport in 3D, to account for effects due to the rotational velocity field and structure.} The grid has a resolution of $100 \times 100$ grid cells with a greater concentration of cells at smaller radii, where line formation typically occurs. The total volume covered by the grid is a sphere of radius $5 \times 10^{17}$ cm. Previous tests from \citetalias{Parkinson2020} indicate that the grid resolution and spherical extent of the wind are sufficient to resolve the line forming region. A complete summary of the grid and parameter values is shown in Table \ref{table: key_parameters}.

\section{Results} \label{sec: results}

    In the following we present the results of our simulations. We first describe the physical characteristics and synthetic spectra for a fiducial model that lies at the center of the parameter space covered by our grid. We then explore how the synthetic spectra depend on the four key parameters covered by our grid (accretion rate, mass-loss rate, terminal velocity and black hole mass).\footnote{The synthetic spectra and \python parameter files for the entire grid are available online at  \hyperlink{https://github.com/saultyevil/tde_optical_reprocessing}{github.com/saultyevil/tde\_optical\_reprocessing} or upon request.}

    \subsection{A fiducial model} \label{section: fiducial_model}
    
        We begin by considering a fiducial "benchmark" model, in which all parameters are set to their central values on the grid: $\text{M}_{\text{BH}} = 3 \times 10^{6}~\text{M}_{\odot}$; $\dot{\text{M}}_{\text{disc}} = 0.15~\dot{\text{M}}_{\text{Edd}}$; $\dot{\text{M}}_{\text{wind}} = 0.3~\dot{\text{M}}_{\text{disc}}$ and $v_{\infty} = 0.3~v_{\text{esc}}$. We now present and describe both the physical properties of the outflow and its optical spectrum.
    
        \subsubsection{Physical properties}
            
        \begin{figure*}
            \centering
            \includegraphics[scale=0.54]{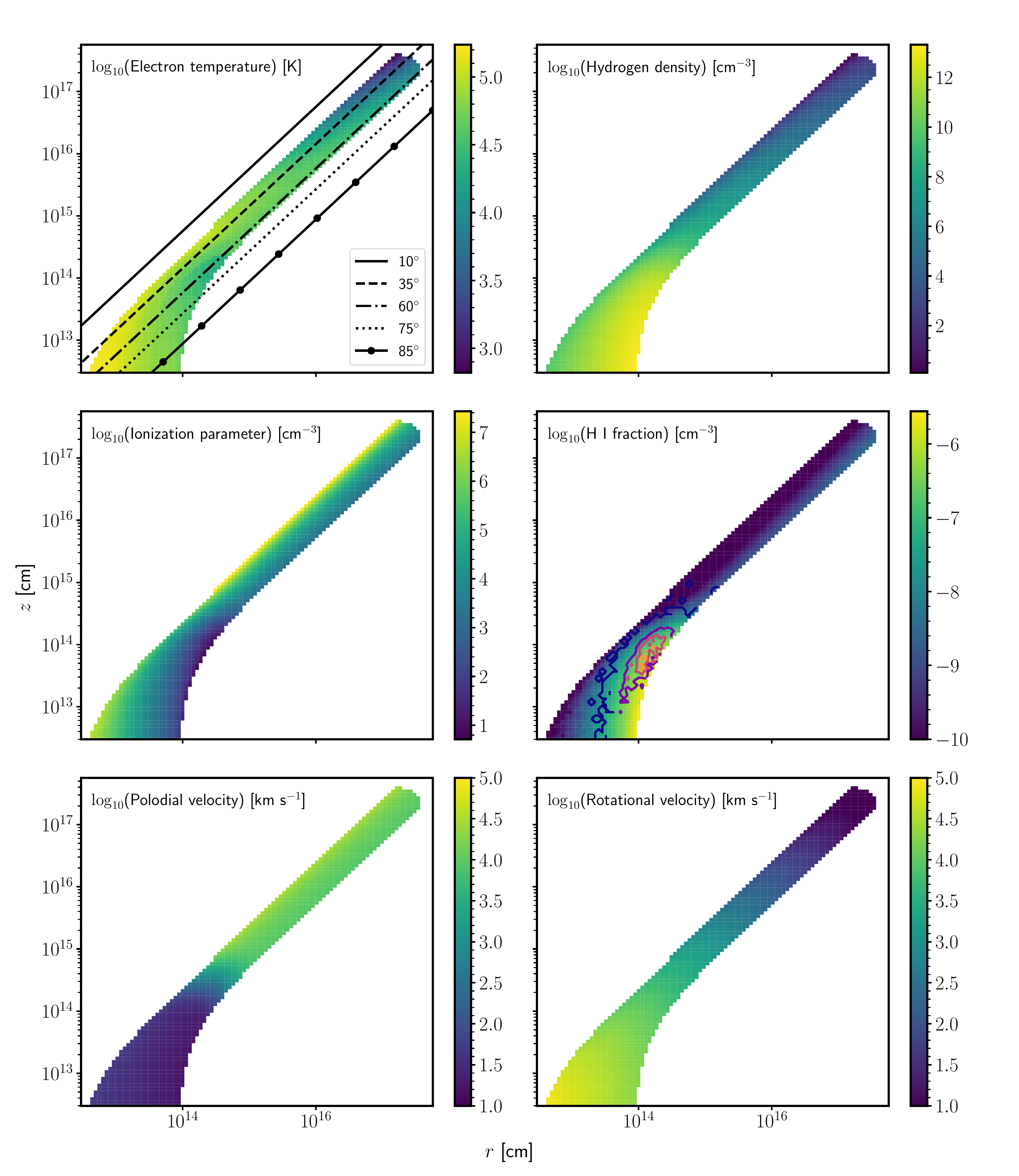}
            \caption{Colour plots showing a selection of physical properties for the fiducial model on logarithmic axes. The $r-z$ plane, where $r$ is the cylindrical radius, is shown and the wind is rotationally symmetric about the $z$-axis. The lines drawn over the wind show sight lines for an observer for inclination angles indicated in the legend. The spatial scales and colour maps are both logarithmic. \textit{Top left:} the electron temperature. {\textit{Top right:} the Hydrogen number density.} \textit{Middle left:} the ionization parameter. {\textit{Middle right:} H \textsc{i} ion fraction.} The contour lines show the origin of \Ha photons contributing to the spectra. From darkest to lightest colour, the contours are regions where the photon count is 0 $N_{\text{tot}}$, 0.25 $N_{\text{tot}}$, 0.5 $N_{\text{tot}}$, 0.75 $N_{\text{tot}}$ and 1.0 $N_{\text{tot}}$ and where $N_{\text{tot}}$ is the total number of \Ha photons. \textit{Bottom left:} the polodial velocity. \textit{Bottom right:} the rotational velocity.}
            \label{fig: fiducial_wind_plot}
        \end{figure*}
 
        In Fig. \ref{fig: fiducial_wind_plot}, we show colour plots for a selection of physical parameters in the fiducial model and five lines showing how and where different sight lines intersect the wind, as well as additional contours showing the relative number density of \Ha photon emission. 
        
        At the base of the wind, the velocity field is rotation dominated and effectively Keplerian, with $v_{\phi} \sim 10^{5}$ \kms~near the inner wind edge and $v_{\phi} \sim 10^{4}$ \kms~at the outer edge. {The launch velocities of streamlines are set as the local sound speed, therefore at the inner and outer edges of the wind the launch velocity is $v_{0} \sim 40~\text{km s}^{-1}$ and $v_{0} \sim 10~\text{km s}^{-1}$, respectively. And by $\sim 10^{15}$ cm, the outflow has accelerated to $10^{4}$ km s$^{-1}$. The velocity at the base of the wind is far lower than that in the \citet{Dai2018} radiation-hydrodynamic model, from which the acceleration length scale is inspired. In this model, the velocity of the outflow is of order $\sim 10^{4}$ km s$^{-1}$ at its base. However, at large radii the velocities are comparable between the two models.}
        
        {Along the disc plane, the matter density increases with radius (due to the choice of $\lambda = 2$) and is $n_{\rm H} \sim 10^{9}$ \pcmc\ at the inner edge  which increases to $n_{\rm H} \sim 10^{12}$ \pcmc\ at the outer edge. At increasing cylindrical radii, both the hydrogen density and rotational velocity decrease. As the electron density is also decreasing, line formation processes which scale with the electron density, such as collisional excitation, also decrease in strength.} The rotational velocity decreases linearly with radius, since material in the outflow is assumed to conserve specific angular momentum. However, the poloidal velocity of the wind increases with radius. The result of this is that Doppler broadening of lines is dominated by rotation close to the disc surface, but by the poloidal velocity field at larger radii. 
        
        At the base of the wind, where it is exposed directly to the radiation generated by the accretion disc, the electron temperature is greatest. The temperature at the very inner edge is in excess of $T_{e} \sim 3 \times 10^{5}$ K, but is closer to $T_{e} \sim 3 \times 10^{4}$ K at the outer edge. At large radii, near the far outer edges of the wind, adiabatic cooling dominates, and the temperature is much cooler ($T_{e} \sim 10^{3}$ K). Since \python does not implement any dust or molecular physics, the treatment of this region of the wind is highly approximate. However, the majority of the line formation we are interested in does not occur in this region. Thus our neglect of this physics should not affect the emergent spectrum to a significant degree.
        
        To measure the ionization state of the wind, we define the ionization parameter 
        \begin{equation}
            U_{\text{H}} = \frac{4\pi}{n_{\text{H}}c} \int_{13.6 \frac{\text{eV}}{h}}^{\infty} \frac{J_{\nu}}{h\nu}~d\nu,
            \label{eq: ion_param}
        \end{equation}
        where $\nu$ denotes frequency, $n_{\text{H}}$ is the number density of hydrogen, $h$ is Planck's constant and $J_{\nu}$ is the monochromatic mean intensity. The ionization parameter measures the ratio of the number density of hydrogen ionizing photons to the number density of hydrogen, making $U_{\text{H}}$ a useful predictor of the global ionization state. However, $U_{\text{H}}$ has no knowledge of the SED shape, meaning it is a poor indicator for the ionization state of ionic species such as C \textsc{iv}.
        
        Fig. \ref{fig: fiducial_wind_plot} shows that the ionization parameter is fairly constant throughout the wind, with {$U_{\text{H}} \sim 10^{5}$ meaning hydrogen is ionized almost everywhere.} At the very top of the wind, the wind is very highly ionized, with {$U_{\text{H}} \sim 10^{7}$}. There is a portion of the wind where {$U_{\text{H}} \lesssim 10$. This part of the wind is located close the the dense base of the wind, at the outer edge. In this region, the wind has been somewhat shielded by the large column of wind material closer in, meaning it is exposed to a softer, reprocessed SED. This results in this part of the wind being cooler. The reduced ionizing flux also means that the neutral hydrogen population is enhanced in this region. This is where the majority of \Ha photons are produced, as shown by the contour lines on the middle right panel of Fig. \ref{fig: fiducial_wind_plot}.}
    
        \subsubsection{Optical spectra}
        
        \begin{figure*}
            \centering
            \includegraphics[scale=0.58]{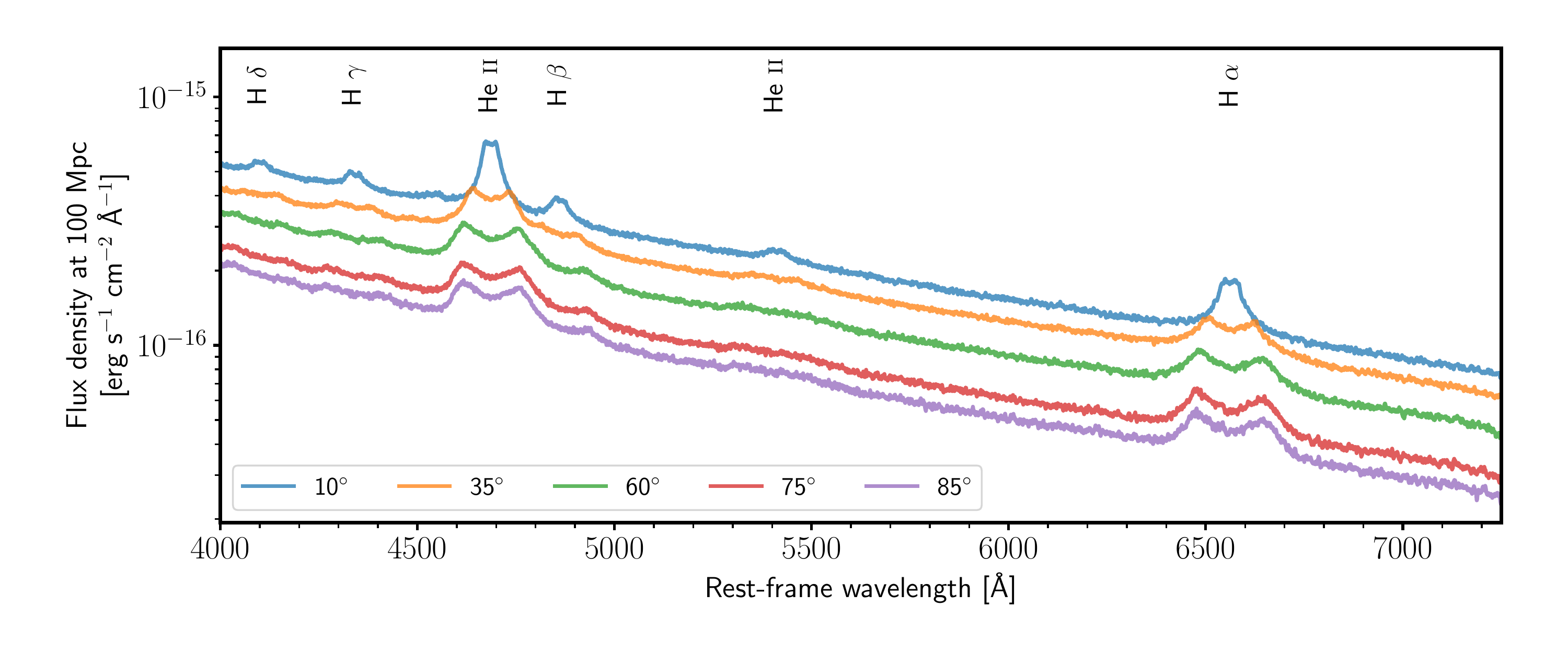}
            \caption{Synthetic optical spectra of the fiducial model. The spectra for five inclination angles are shown, each coloured as labelled by the legend. For each inclination angle, the model produces broad Balmer and helium recombination features. At low inclinations, the emission lines are single peaked but transition into double peaked lines as the inclination angle increases. Notably, the intermediate and high inclination spectra are almost identical as to the base of the wind, which these sight lines cut through, dominates spectrum generation. The spectra are plotted on a log-linear axes and have not been smoothed. Labelled at the top of each panel are important line transitions.}
            \label{fig: fiducial_optical_spectra}
        \end{figure*}
        
        The synthetic optical spectra produced by the fiducial model for five sight lines are shown in Fig. \ref{fig: fiducial_optical_spectra}. {For each inclination angle, the model produces broad, double peaked, recombination emission in \Ha and \Hbstop. The spectra also include other \atomictransition{He}{ii} recombination lines commonly seen in the spectra of TDEs. The \atomictransition{He}{ii} emission line is clearest at low inclination; it becomes very broad and blended with \Hb for intermediate and edge-on inclinations, as these probe higher projected velocities.}

        {The line widths and peak-to-peak separations of the double-peaked emission lines increase with inclination. This happens because rotation is the dominant line broadening mechanism, and edge-on sightlines see higher projected rotational velocities. This also explains the double-peaked nature of the emission lines. Rotational kinematics dominate in the line-forming regions. The emissivity of  recombination lines scales with density squared, so these features are formed preferentially in the dense base of the wind where the poloidal velocities are still relatively low. In fact, the lines are mostly formed near the outer edge of the wind base, where the outflow is both dense and relatively more neutral (since it is exposed to the redder, reprocessed SED, instead of the ionizing disc SED).}

        The emission lines are asymmetric, displaying an extended red wing. TDEs, such as ASASSN-14ae \citep{Holoien2014}, AT2019qiz \citep{hung_discovery_2020} or AT2018zr \citep{holoien_ps18kh_2019}, often exhibit the same asymmetry, deviating from a Gaussian line profile. \citet{roth_what_2018} proposes that the origin of this asymmetry is from line formation in an optically thick outflow. Photons which scatter in an optically thick expanding outflow lose energy through adiabatic work on the outflow, redshifting their frequency resulting in an extended red wing in emission lines \citep[e.g.][]{laurent_effects_2007, Strubbe2009, Metzger2016a, roth_what_2018}, in a process \citet{roth_what_2018} coins as \textit{adiabatic reprocessing}. Other than through energy losses due to adiabatic work, extended red wings could instead form due to (general) relativistic effects \citep[e.g.][]{ERACLEOUS2009133}.
        
        \begin{table} 
            \centering
    		 \resizebox{\columnwidth}{!}{
        		 \begin{tabular}{cccccc}
	                \hline
        			 Inclination  & Column density        & \multicolumn{4}{c}{Optical depths}                                      \\
        			 ($^{\circ})$ & (cm$^{-2}$)           &       &                          &            &                        \\
        			 \hline
        			              & Hydrogen              & Lyman & \atomictransition{He}{i} & \atomictransition{He}{ii} & Electron\\
        			 \hline
        			 35           & $1.02 \times 10^{25}$ & $1.13 \times 10^{-3}$ & $2.4 \times 10^{-4}$ & $4.00 \times 10^{-4}$ & 0.82\\
        			 60           & $2.46 \times 10^{26}$ & 2.75                  & 0.58                 & $4.01 \times 10^{3}$ & 19.8\\
        			 75           & $8.19 \times 10^{26}$ & 190                   & 5.5                  & $3.41 \times 10^{5}$ & 56.2\\
        			 85           & $8.42 \times 10^{26}$ & 5.49                  & 1.17                 & $1.48 \times 10^{4}$ & 54.2\\
        			 \hline
        		 \end{tabular}
     		 }
    		 \caption{{The hydrogen column density and the integrated optical depth for three frequencies corresponding to three photo-ionization edges and electron scattering for the fiducial model in Section \ref{section: fiducial_model}. In cases where the optical depth are constant across multiple frequencies, the electron scattering opacity is dominating the opacity from photo-absorption.}}
            \label{table: fiducial_opacity_parameters}
        \end{table}
        
        The spectra produced by our fiducial model {at high inclinations are very similar}, with similar line strengths and widths. These sight lines intersect the dense base of the wind, and are probing parts of the wind which have very similar conditions. In Table \ref{table: fiducial_opacity_parameters}, we tabulate the hydrogen column density and {the integrated optical depths for three photo-ionization edges} along four of the sight lines shown in the top left panel of Fig. \ref{fig: fiducial_wind_plot}. Along these sight lines, the integrated optical depths and column densities are very high ($\text{N}_{\text{H}} \sim 10^{26}$ cm$^{-2}$). They are so high, in fact, that this region of the wind actually dominates the formation of the optical spectrum. 
        
        As an example, for inclinations close to the disc plane, {the optical depth of the \atomictransition{He}{ii} edge is around $\tau \sim 10^{4} - 10^{5}$}, suggesting that photo-ionization of helium ions in the base of the wind is contributing significantly to reprocessing the disc emission. {In fact, this is the main reprocessing mechanism in our outflows}. Electron scattering optical depths are also fairly large for these inclinations, with {$\tau \simeq 55$}. By contrast, at low inclinations, the electron scattering optical depth is only {$\tau \lesssim 1$}, and the optical depths associated with the photo-ionization edges are broadly comparable. Even so, these sight lines can still intersect a small fraction of the dense wind base and thus see a substantial column density (e.g. $\text{N}_{\text{H}} \sim 10^{25}$ cm$^{-2}$ for $i=35^{\circ}$). Electron scattering is the dominant global source of opacity along low inclination sight lines, since the reduction in photo-ionization opacity means electron scattering reprocesses across a wider frequency range. 

    \subsection{Synthetic spectra across the simulation grid} \label{section: spectra_results}
    
        \begin{figure*}
            \centering q
            \includegraphics[scale=0.525]{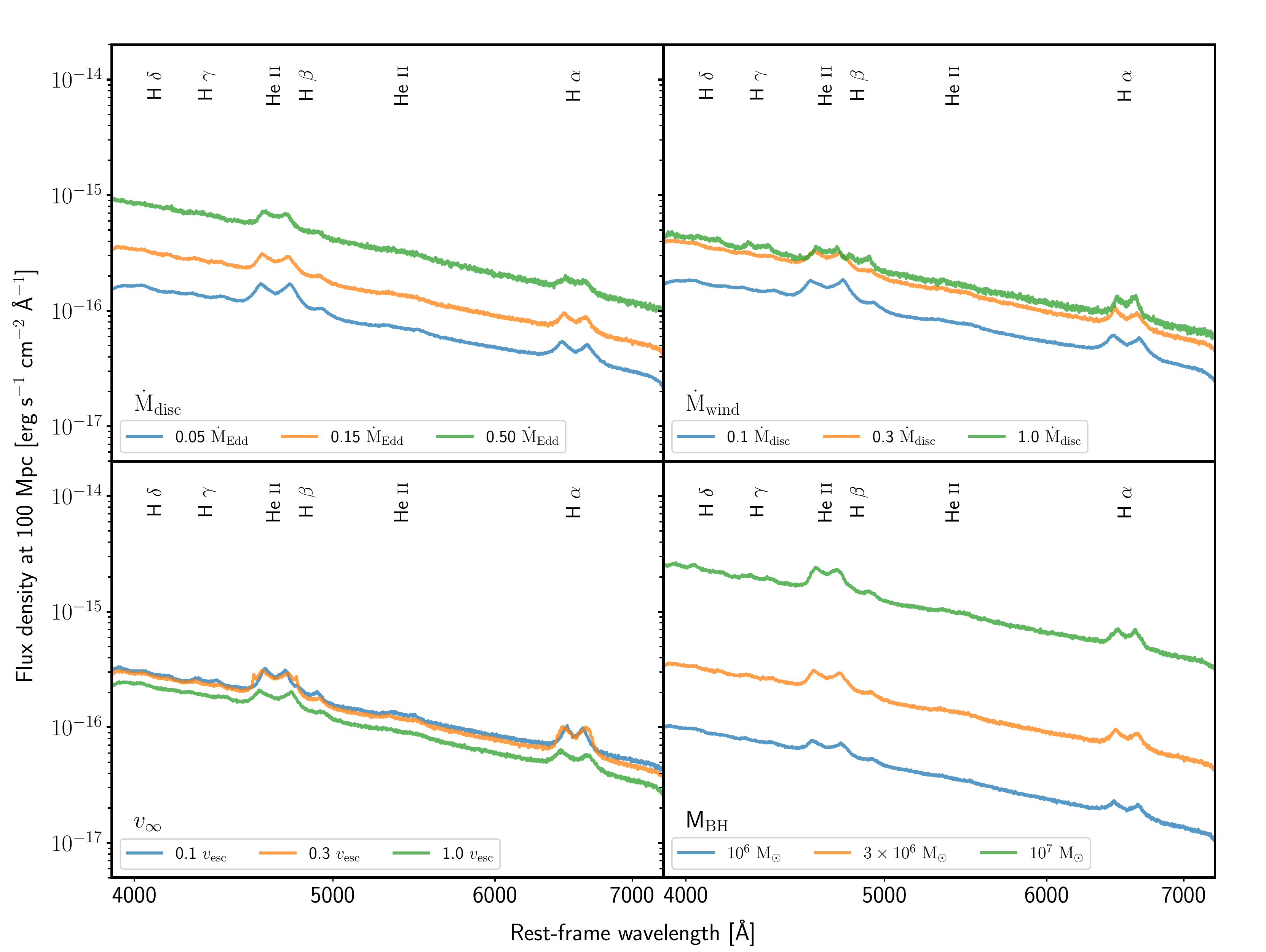}
            \caption{Synthetic optical spectra for the four parameter grids on log-linear axes. The first three panels show spectra for the mass accretion rate ($\dot{\text{M}}_{\text{disc}}$), the wind mass-loss rate ($\dot{\text{M}}_{\text{wind}}$) and the terminal velocity of wind streamlines ($v_{\infty}$) grids, using a black hole mass of $\text{M}_{\text{BH}} = 3 \times 10^{6}~\text{M}_{\odot}$. The bottom right panel shows spectra of the fiducial model for three black hole masses ($\text{M}_{\text{BH}}$). In each panel, the orange spectrum corresponds to the fiducial model in Section \ref{section: fiducial_model}. All of the spectra exhibit broad Balmer and helium recombination features, strongly reminiscent of TDE-Bowen objects. The emission lines are double peaked and feature an extended red wing. Each spectrum is for an intermediate, in-wind, inclination angle of $60^{\circ}$. The spectra have not been smoothed. Labelled at the top of each panel are important line transitions.}
            \label{fig: optical_line_spectra}
        \end{figure*}
 
        In Fig. \ref{fig: optical_line_spectra}, we show an intermediate-inclination spectrum for each model in our simulation grid (as labelled in the legend). Note that the viewing angle adopted for these plots, $i = 60^{\circ}$, corresponds to a sight line which looks into the wind cone. Additionally, in Fig. \ref{fig:  fiducial_mass_reprocessing}, we show angle-integrated disc and emergent reprocessed SEDs for the same models, along with the continuum optical depths they present along several sight lines.
        
        \begin{figure*}
            \centering
            \includegraphics[scale=0.53]{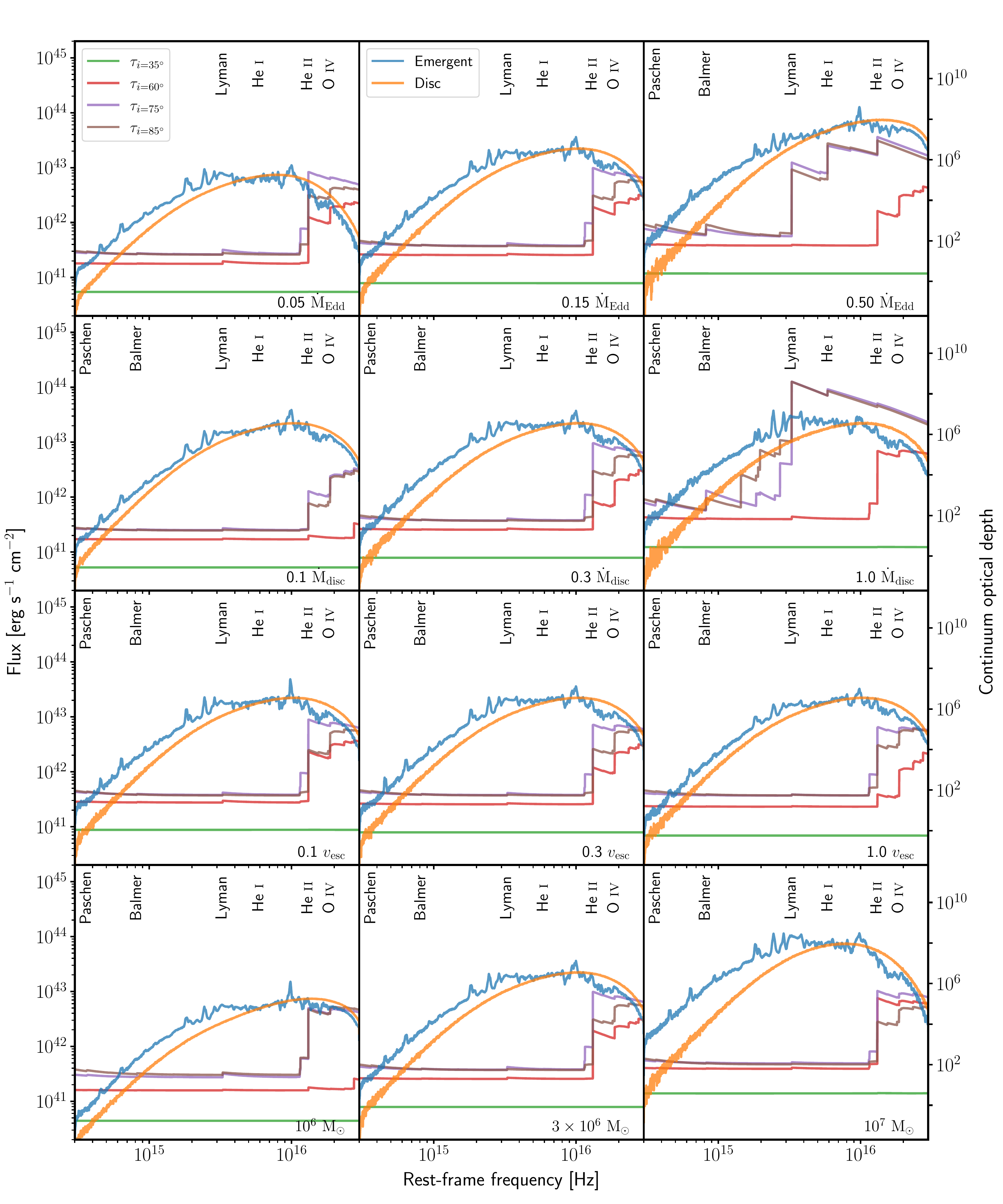}
            \caption{The continuum optical depth as a function of frequency for various sight lines (right axis) and both the emergent and disc SEDs for the same models as in Fig. \ref{fig: optical_line_spectra}. The middle column corresponds to the fiducial model from Section \ref{section: fiducial_model}. The spectra have been plotted on logarithmic axes and important photo-ionization edges have been labelled at the top. Each model is producing a redder (reprocessed) SED relative to the input \textit{disc} SED. In general, denser winds are more efficiently reprocessing the disc SED and result in optically bright spectra. The main reprocessing mechanism is photo-ionization of \atomictransition{He}{i} and \atomictransition{He}{ii}, but in some winds H \textsc{i} also contributes. The electron scattering optical depth (corresponding to the ``flat'' region of the optical depth spectra) is roughly the same between models.}
            \label{fig: fiducial_mass_reprocessing} 
        \end{figure*}
        
        For all of the models, the optical spectra feature broad Balmer and helium recombination lines that are strongly reminiscent of the TDE-Bowen class of spectra.\footnote{Note that \python does not currently include the Bowen fluorescence process, so the models cannot produce the Bowen features themselves.}  {However, the \atomictransition{He}{ii} and \Hb emission lines can often become blended with one another, making them difficult to distinguish from each other.} 
        
        All the synthetic spectra shown in Fig. \ref{fig: optical_line_spectra} feature double-peaked lines and asymmetric red-skewed line wings, for the same reasons as the fiducial model discussed earlier in Section \ref{section: fiducial_model}. No model on our simulation grid creates the pure Balmer recombination spectra (without helium lines) that are characteristic of the TDE-H class. We suspect this is a signature of the radiation field in the optical line-forming regions. More specifically, the absence of helium recombination features in TDE-H systems suggests a lack of photons with energies sufficient to ionize helium (24.6 eV). By contrast, all of our models contain regions whose ionization state and emission measure are both high enough to create observable \atomictransition{He}{i} and \atomictransition{He}{ii} features.

        In any case, Fig. \ref{fig: fiducial_mass_reprocessing} shows that each of our wind models, regardless of its detailed parameters, strongly reprocesses the accretion disc continuum (labelled as "disc"),  producing a much redder observed SED (labelled as "emergent"). Thus reprocessing of the disc continuum in an outflow can, rather naturally, result in an optically bright emergent SED.
        
        Qualitatively, the line spectra and SEDs do not change drastically with different parameter values or with inclination. More specifically, all the synthetic line spectra produce broadly the same set of transitions, accompanied by a continuum with roughly similar colour temperatures. The ionization state of the wind, in particular that of hydrogen and helium, is insensitive to moderate changes in the input SED, the mass-loading of the wind or its kinematics. {The dense base of the wind, which is the main line-forming region, does not change all that much between models.} This suggests that reprocessing, and the production of Balmer and helium recombination features, can happen over a fairly broad range of conditions in TDEs. Quantitatively, the spectra do, of course, depend on the model parameters. In the following sections, we outline the main effect of the different parameters we have explored on the resulting line spectra and emergent SEDs.
    
        \subsubsection{Disc mass-accretion rate}
        
        The top left panel of Fig. \ref{fig: optical_line_spectra} shows three models with different disc accretion rates. In all of these models, the wind mass-loss rate is 30\% of the accretion rate. The main difference between the spectra is the continuum normalization and the strength of the \Ha line. The highest accretion rate model produces the most luminous spectrum. {The increase in the continuum level is mostly due to a more luminous disc, but also because of greater reprocessing associated with the higher density of the wind. This is clear by looking at the top row of Fig. \ref{fig: fiducial_mass_reprocessing}, which shows the reprocessed SED relative to the input SED for the three accretion rates. In the highest accretion rate model, the high frequency region of the disc continuum is slightly less attenuated, but there is significant optical depth from lower frequency edges contributing to reprocessing the SED. So while the higher continuum luminosity can mostly be attributed with the higher accretion rate, it does appear this wind reprocesses more radiation and re-emits it at longer wavelengths.}
        
        {
        The spectrum produced with the lowest accretion rate in Fig. \ref{fig: fiducial_optical_spectra} has the strongest emission lines. In general, the strength of the emission lines decreases with accretion rate, even though the higher accretion rate models are reprocessing (and re-emitting via recombination) more emission and that the emissivity of these lines scales with the density squared. In this situation, the lines appear weaker because of the stronger optical continuum due to stronger reprocessing decreasing the line-to-continuum contrast.
        }

        \subsubsection{Wind mass-loss rate}
        
        The top right panel in Fig. \ref{fig: optical_line_spectra} shows how changing the mass-loss rate of the wind affects the line spectrum, and the second row of Fig. \ref{fig: fiducial_mass_reprocessing} shows how the reprocessed SED changes. The model featuring the highest mass-loss rate (and thus highest densities) has the strongest recombination features, since, stating this once again, the emissivity of recombination scales with the density squared. This model also produces the brightest optical spectrum, since it reprocesses the largest fraction of the disc emission, although the middle mass-loss model is not far behind.
        
        Quite generally, denser winds tend to be more efficient at reprocessing disc emission, generating optically brighter spectra. This derives from the dominant reprocessing mechanisms being photo-ionization and free-free absorption, and since the optical depth of edges, specifically \atomictransition{He}{ii}, and the electron/ion density, which free-free opacity scales with, are enhanced in a denser wind, the reprocessing becomes stronger. In the lowest mass-loss rate model, the input SED is reprocessed less (refer to Fig. \ref{fig: fiducial_mass_reprocessing}) {and the emergent SED is the least red. The opacity of the Lyman and \atomictransition{He}{i} photo-ionization edges are considerably diminished and comparable to the scattering optical depth, resulting in a reduction in the overall amount of reprocessing via photo-ionization. However, as the \atomictransition{He}{ii} edge is still comparably strong, there is still enough reprocessing via photo-absorption to create an optically enhanced spectrum. It appears that the \atomictransition{He}{ii} edge is very important in reprocessing high frequency disc emission.}
        
        \subsubsection{Terminal velocity}
        
        The terminal velocity is one of the parameters which controls the wind kinematics. Larger terminal velocities imply higher velocities \textit{along} streamlines, as well as lower densities (via the continuity equation). Thus slower outflows are denser and produce redder reprocessed SEDs (third row down of Fig. \ref{fig: fiducial_mass_reprocessing}). However, the effect of the terminal velocity on the optical continuum luminosity and colour temperature is relatively modest. 
        
        The bottom left panel of Fig. \ref{fig: optical_line_spectra} shows the synthetic spectra for the three different values of $v_{\infty}$ on our grid. Since the density is modified by the velocity, {the two models with the lowest terminal velocities, i.e. which have largest densities at the base of the wind, produce the strongest recombination lines and the brightest optical continuum. Much like changing the mass-loss rate of the wind, varying $v_{\infty}$ from $0.1-1 ~ v_{\rm esc}$ does not significantly affect the overall optical spectrum, producing only modest changes in the normalisation. However, as noted previously, changing the acceleration length scale has a more significant impact.}
        
        \subsubsection{Black hole mass}
        
        The bottom right panel of Fig. \ref{fig: optical_line_spectra} shows the optical spectra produced by the fiducial wind model for three different black hole masses. Changing the black hole mass indirectly affects other parameters of the models as well, such as the three other parameters comprising our grid (since these parameters all scale with black hole mass). 
        
        {
        The most significant differences between the spectra, associated with different black hole masses, is their continuum luminosities and their line strengths. The model with the most massive black hole produces the strongest lines, as expected since recombination scales with the density squared.  The same model produces the brightest optical spectrum, in this case because the accretion disc is the brightest in the grid, rather than because of any significantly greater level of reprocessing relative to the rest of the grid. In the case of the highest black hole mass model, there are simple ``more'' photons to reprocess. Fig. \ref{fig: optical_line_spectra} also shows that the same Balmer and helium recombination features are produced for all three black hole masses. Otherwise, the spectra remain relatively unchanged in terms of distinct spectral features.
        }

\section{Discussion} \label{sec: discussion}

    \subsection{The origin of optical emission}

        In early TDE models, the stellar debris is assumed to fall back onto the black hole, typically, at a super-Eddington rates. The luminosity of these events are then generated via thermal blackbody radiation from a hot, qausi-circular accretion disc \citep[see, e.g.,][]{Rees1988}. The prevalence of optically bright TDEs with a weak soft X-ray component is, then, somewhat inconsistent with these early models. Although the early models of TDEs assumed that the radiation observed would be that observed directly from the hot accretion disc, most modern interpretations of these events involve a significant amount of radiative reprocessing to explain the optical brightness of some TDEs.
        
        {Reprocessing is a core component in a number of TDE models \citep{Strubbe2009, Metzger2016a, Roth2016, Dai_2008, roth_what_2018, Lu_2020, Piro_2020, bonnerot_2020}.} Notably, \citet{Roth2016} showed that reprocessing in an optically thick outflow plays a key role in the formation of the optical spectrum. They identified two distinct reprocessing regimes, which can explain why some TDEs can still retain their soft X-ray component. Reprocessing also plays a key role in the unification scenario proposed by \citet{Dai2018}. In their model, a given TDE could be observed as either optically or X-ray bright, depending on the inclination of the observer. Intermediate and high-inclination observers see an optically bright SED, since the disc emission is reprocessed via adiabatic cooling and photo-ionization along these sight lines. By contrast, observers viewing the TDE face-on see an X-ray bright TDE, since they are looking directly at the exposed inner accretion disc.
        
        In contrast to the dense, spherically symmetric model of \citet{Roth2016}, our biconical disc wind models typically do not not completely absorb the soft X-ray flux. This difference is likely a geometric effect, as photons are able to escape along paths with lower optical depths in our outflows. Thus, similar to \citet{Dai2018}, we find that the ratio of optical to X-ray fluxes increases with inclination angle. At polar angles, the SED is more akin to a underlying disc spectrum, since the optical depth is lower -- and reprocessing less efficient -- along these sight lines.
        
        In Section \ref{sec: results}, we show how reprocessing in an optically thick accretion disc wind, with plausible parameters, results in an optically bright spectrum, significantly redder than the original accretion disc SED. The main reprocessing mechanism in our winds is photo-ionization, which is also the case for the \citet{Dai2018} unification model. {Even so, free-free absorption is \textit{still} important with an optical depth of $\tau_{\rm ff} \sim 10^{2}$ for intermediate and high inclinations in the fiducial model.} The majority of the photo-ionization reprocessing is associated with the ionization of \atomictransition{He}{ii}; in some models, \atomictransition{He}{i} and H \textsc{i} are also important (Fig. \ref{fig: fiducial_mass_reprocessing}). When the opacity in both helium edges is significant, high frequency radiation is efficiently absorbed and reprocessed, softening the SED and reducing the EUV/X-ray brightness. The absorbed luminosity is re-emitted at longer wavelengths via recombination, free-free and line emission. 
        
        {
        The photo-ionization of \atomictransition{He}{ii} seems to be particularly important for the reprocessing efficiency. For instance, the optical brightness in the wind mass-loss rate sub-grid (second row down in Fig. \ref{fig: fiducial_mass_reprocessing}) increases as the optical depth (and hence absorption) of this edge increases, as is clear by comparing the $0.1~\dot{\rm M}_{\rm disc}$ and $0.3~\dot{\rm M}_{\rm disc}$ models. This is, in-fact, a common trend across all of the models in Fig. \ref{fig: fiducial_mass_reprocessing}. But additionally, models with high optical depths for lower frequency edges (e.g. \atomictransition{He}{i} and Lyman) also have optically brighter spectra (see the top right panel of the first and second row in Fig. \ref{fig: fiducial_mass_reprocessing}). It seems that reprocessing of not just the soft X-ray photons, but also of the EUV photons is important for the production of optically bright spectra.
        }
        
 		{
        As mentioned previously, photons undergoing repeated scattering events in a divergent flow lose energy and are redshifted as they do work on the outflow \citep[see, e.g.,][]{laurent_effects_2007, roth_what_2018}. This effect, sometimes referred to as ``adiabatic reprocessing'', can be important \citep[see Fig. 14 in][]{roth_what_2018}. The integrated electron scattering optical depth in our winds are significant ($\tau \gtrsim 50$ for intermediate and high inclinations in the fiducial model), and photons are almost certain to undergo multiple scatters before reaching an observer. However, compared to outflow models in which adiabatic reprocessing is known to dominate, the scattering optical depths in our winds are fairly modest. For example, \citet{roth_what_2018} report a scattering optical depth of $\tau \simeq 130$ for their spherical outflow, in which repeated scatterings in the outflow play a major role in redshifting the SED. Additionally, as the models here are 2D, photons are able to scatter around highly optically thick regions and do not become trapped, as they would in a spherical model such as by \citet{roth_what_2018}.
        }
 
        {
        In our fiducial model, the role played by adiabatic reprocessing is minor. The scattering optical depths are similar between the models in Fig. \ref{fig: fiducial_mass_reprocessing}, but the amount of reprocessing is not. It is clear, then, that whilst adiabatic reprocessing can modify the SED, in our models it is not an important process for the creation optically bright spectra. To see this, consider the $\dot{\text{M}}_{\text{wind}} = 0.1~\dot{\text{M}}_{\text{disc}}$ panel in Fig. \ref{fig: fiducial_mass_reprocessing}. Here, the scattering optical depth is similar to other models, but the optical depth of the \atomictransition{He}{i} edge is greatly reduced and comparable to the scattering optical depth. The optical luminosity in this model is low, showing that it tracks the photo-ionization optical depth, rather than the scattering one. Therefore, adiabatic reprocessing is not a dominant effect in our outflow models.
        }
        
        \begin{figure*}
            \centering
            \includegraphics[scale=0.55]{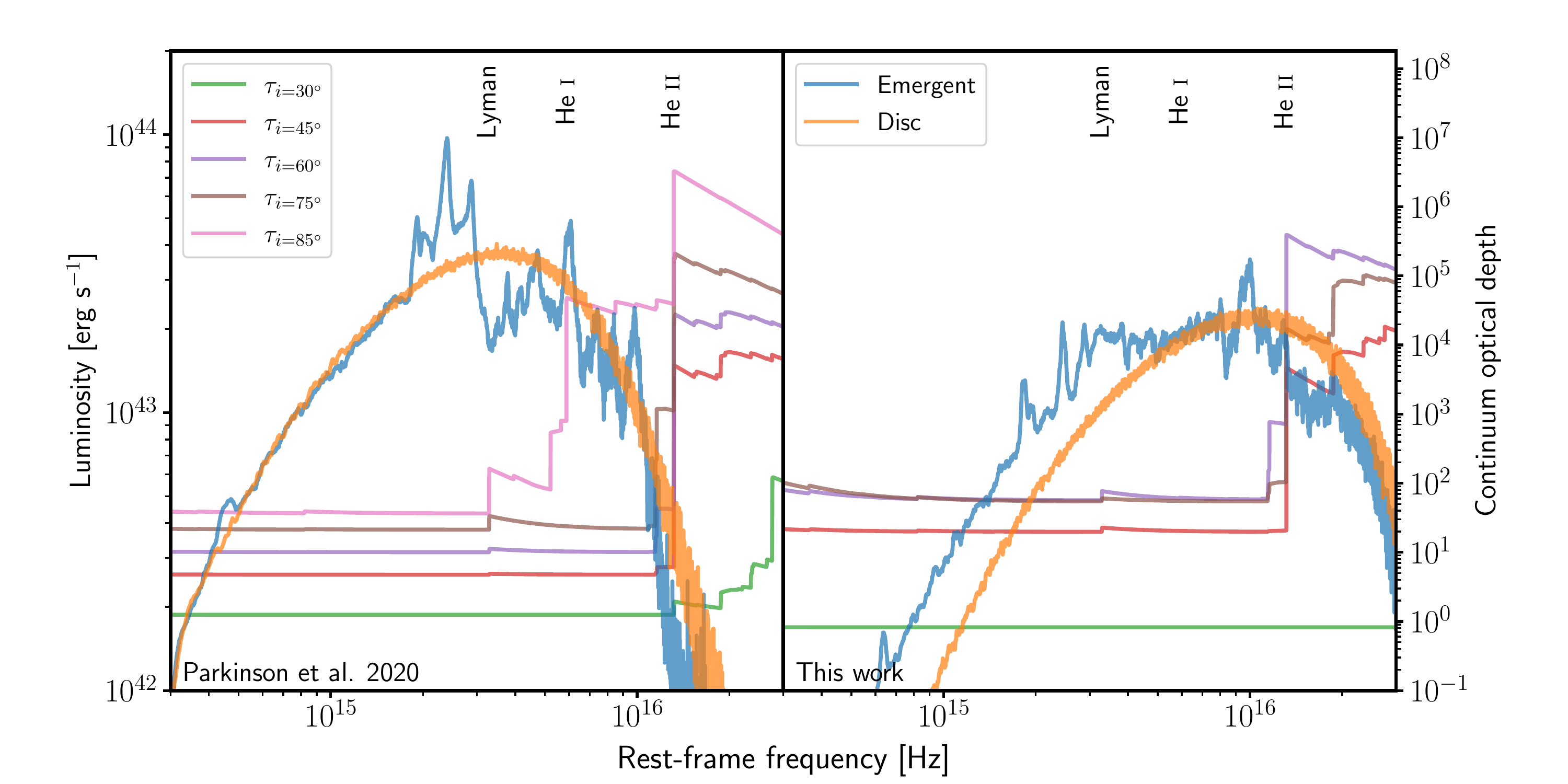}
            \caption{The continuum optical depth as a function of frequency for various sight lines (right axis) and both the emergent and disc SEDs (left axis) for the fiducial model in \citetalias{Parkinson2020} and and in this work (Section \ref{section: fiducial_model}). Both axes are logarithmically spaced and important photo-ionization edges have been labelled at the top. Relative to the \citetalias{Parkinson2020} fiducial model, reprocessing is more efficient in the fiducial model presented in this work. Crucially whilst the both models have similar optical depth for the same photo-ionization edges, the disc continuum peaks at a higher frequency in the ``new'' fiducial model. In the \citetalias{Parkinson2020} model, the disc SED does not include enough \textit{ionizing} photons resulting in reprocessing via photo-ionization becoming inefficient.}
            \label{fig: old_new_reprocessing} 
        \end{figure*}
        
        The emergent optical luminosity is also affected by the SED of the accretion disc. If the SED does not have enough high frequency flux, then there is nothing to reprocess to create a red optical spectrum. This effect is illustrated in Fig. \ref{fig: old_new_reprocessing}, with the fiducial model we used in \citetalias{Parkinson2020} and the fiducial model used in the present work. Clearly, the new fiducial model reprocesses a great deal more of the underlying disc continuum. This is despite the fact that the photo-ionization optical depth for \atomictransition{He}{ii} is not too dissimilar between the models. {But there is a disparity between the optical depths for the \atomictransition{He}{i} and Lyman edges in the old fiducial model, where they are larger.}
        
        So what is responsible for the higher reprocessing efficiency of our new fiducial model, compared to our \citetalias{Parkinson2020} model? Crucially, the disc continuum in this model peaks at a higher frequency, which means a higher fraction of its luminosity is susceptible to reprocessing by the \atomictransition{He}{ii} edge. The reprocessing efficiency is not just related to the physical properties of the wind, notably its (column) density, but it also depends on the underlying SED. If this SED does not include enough \textit{ionizing} photons, reprocessing via photo-ionization becomes inefficient (as does free-free reprocessing). Of course, adiabatic scattering can still reprocess the continuum if the outflow is optically thick, even if this is a relatively small effect in our models.
            
    \subsection{Formation of optical lines} \label{sec: formation_of_optical_lines}
    
    {
    When the density of the wind is low, the Balmer and \atomictransition{He}{ii} optical features become weak. If the density is too low to shield or soften the disc emission via reprocessing, the recombination features are greatly diminished as the wind becomes overionized, preventing the formation of absorption and emission lines \citep[e.g.][]{Proga2000, Proga2002, Higginbottom2013}. More generally, however, the strength of recombination features scales with emission measure, $\text{EM} = \int n_{e}^{2}~dV$. This depends not only the local density, but also on the emitting volume, so where in the outflow these lines form is also important. For example, emission lines will preferentially form near the base of the wind where the density is largest. But, if \Ha photons were being emitted over an extended region in the sparse, outer portion of the outflow, this emission could still be significant, since the volume of this region is so large.}
    
    {
    One way to modify the density of the outflow in our kinematic approach (other than by modifying the mass loss rate) is by changing the acceleration length scale or the acceleration exponent. Together, these variables control how quickly the outflow accelerates. By increasing the acceleration length and increasing the exponent, the outflow density increases and the ionization state is lowered. Both effects result in stronger recombination features, so the appearance of the optical spectra is sensitive to these kinematic model parameters. However, some lines, such as \Hastop, are insensitive to changes in the velocity structure. This suggests these line photons are produced in the very base of the wind, whose kinematic properties remain fairly constant with the change in parameters.}

    \subsection{The effect of electron scattering on the line profiles}
    
        \begin{figure*}
            \centering
            \includegraphics[scale=0.55]{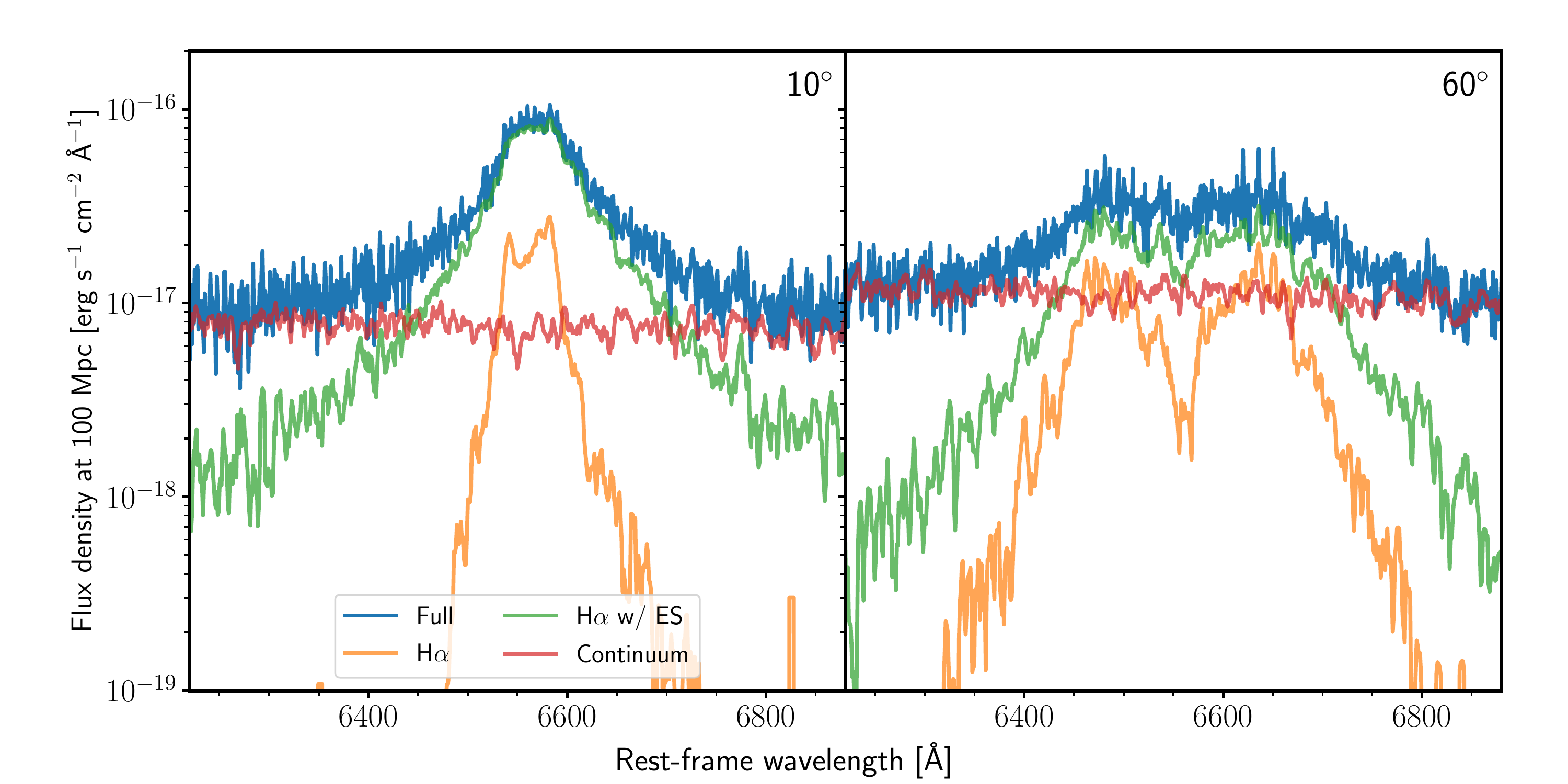}
            \caption{Synthetic spectra for two inclinations (labelled in the top right of each panel) of photons contributing toward the emergent spectrum (labelled as full as in the figure) of the \Ha line profile for the fiducial model. Included are ``pure'' continuum photons (free-free, bound-free and electron scattered), as well as \Ha photons and \Ha photons which have electron scattered at least once between emission and escape (\Ha w/ ES). The spectra have been smoothed by a six pixel boxcar filter. Importantly, electron scattering is responsible for the broad wings of the emission lines. It also smooths out the double peaked line profile at low inclinations. \textit{Left:} a low inclination spectrum (10\degrees). \textit{Right:} an intermediate inclination spectrum (60\degrees).}
            \label{fig: spec_breakdown}
        \end{figure*}
    
        In Fig. \ref{fig: spec_breakdown}, we break down the contribution of different physical processes on the spectrum produced by our fiducial model in the vicinity of \Hastop, isolating the contribution of ``continuum'' photons which have never undergone any bound-bound interaction in the outflow. This category includes photons emitted directly by the disc and photons produced via bound-free or free-free processes. We also isolate ``line'' photons which have undergone one or more electron scatters since their bound-bound interaction (\Ha w/ES). The final contribution is "pure" \Ha photons, i.e. photons whose last interaction in the wind was in the bound-bound \Ha transition.
        
        Fig. \ref{fig: spec_breakdown} shows very clearly that electron scattering is responsible for the broad wings of the emission line, in line with the results of \citet{roth_what_2018}. In their simulations, higher electron scattering optical depths resulted in broader emission lines. In our models, electron scattering can also smooth out the double peaked line profile, at least at low inclinations (left panel). 
        
    \subsection{Comparison to observations}
    
\begin{figure*}
            \centering
            \includegraphics[scale=0.55]{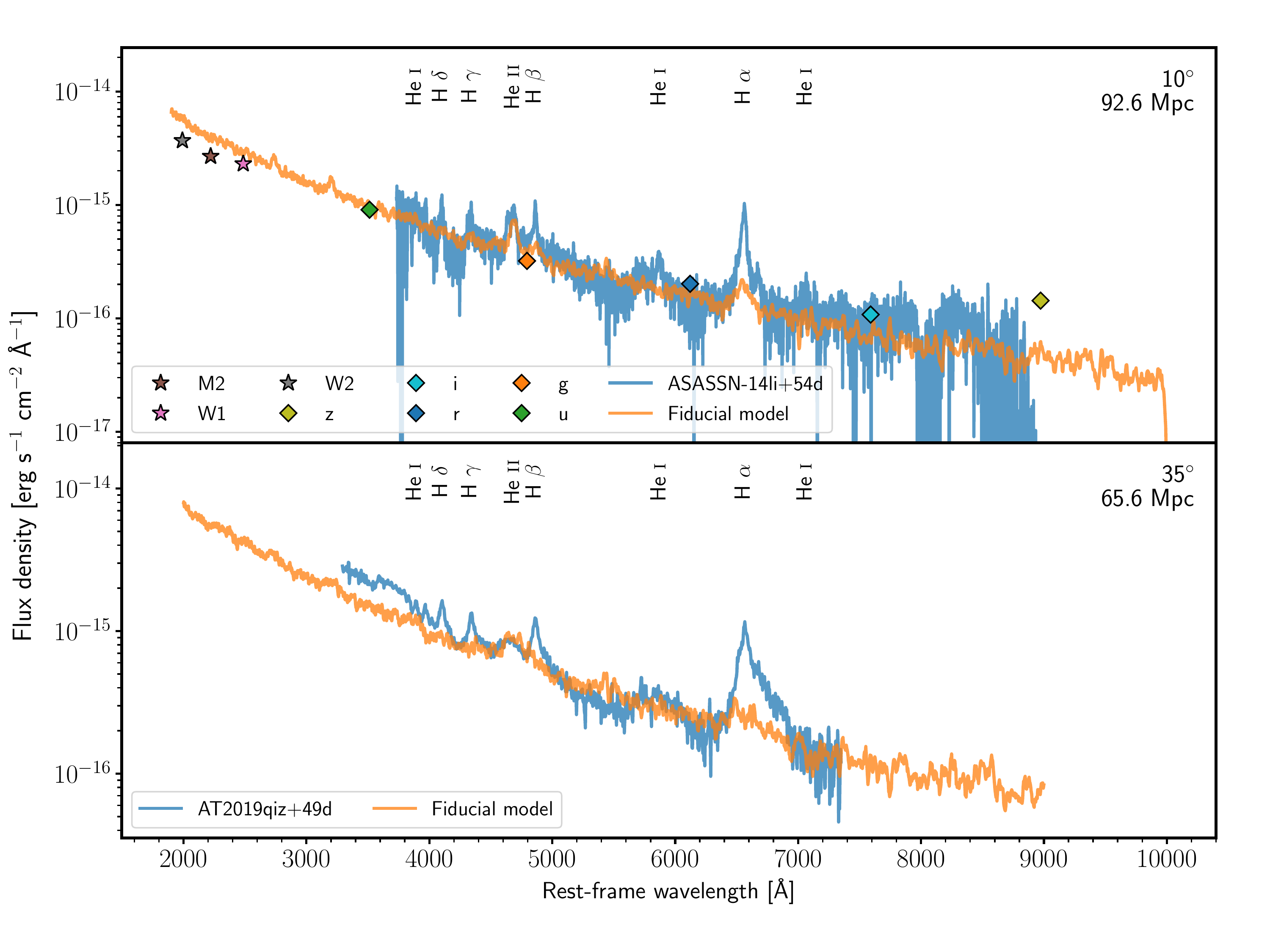}
            \caption{Rest frame synthetic optical spectra for two models (labelled in the legend) plotted on linear axes and against the optical spectra of ASASSN14-li bat $\Delta t = 54$ days \citep{Holoien2016} and AT2019qiz at $\Delta t = 49$ days \citep{Nicholl2020}. The spectra for the objects have been corrected for foreground extinction assuming a \citet{Cardelli1989} extinction curve with $R_{v} = 3.1$ for both TDEs and $E(B-V) = 0.0225$ and $E(B-V) = 0.0939$ for ASASSN14-li and AT2019qiz respectively. {Photometry for ASASSN-14li, taken from \citet{Holoien2016}, has also been included with the filters labelled in the legend.} The distance of the object and inclination of the model are labelled in the top right of each panel. The spectra have not been smoothed. Important line transitions are marked at the top of each panel. The synthetic spectra are in rather excellent agreement with the observations, especially considering neither model has been tailored to fit the specific observation, producing the same recombination features with similar continua.} 
            \label{fig: tde_model_comparison}
        \end{figure*}

        In Fig. \ref{fig: tde_model_comparison}, we show comparisons of two of our models to the optical spectra of ASASSN14-li and AT2019qiz. Other than the choice of inclination angles and re-scaling the spectra to the relevant distance, the model parameters used to generate the synthetic spectra have not been tailored and/or fine-tuned to fit either object. The observed spectra were taken at roughly the same time post peak, at $ \Delta t \approx 50$ days, and have been corrected for foreground extinction assuming a \citet{Cardelli1989} extinction curve with $R_{v} = 3.1$ with $E(B-V) = 0.0225$ and $E(B-V) = 0.0939$ for ASASSN14-li and AT2019qiz, respectively. {For ASASSN-14li, we have also included photometry, provided by \citet{Holoien2016}, from both the \textit{Swift} UltraViolet and Optical Telescope \citep{roming_swift_2005} and from the 2-m Liverpool telescope \citep{steele2004proc}. For the comparison with ASASSN-14li, we use a black hole with mass $\text{M}_{\rm BH} = 3 \times 10^{6}~\text{M}_{\odot}$ and a mass loss rate of $\dot{\rm M}_{\rm wind} = \dot{\rm M}_{\rm disc}$. For AT2019qiz we use a model with $\dot{\rm M}_{\rm disc} = 0.5~\dot{\rm M}_{\rm Edd}$ and a black hole mass $\text{M}_{\rm BH} = 10^{6}~\text{M}_{\odot}$. The rest of the model parameters are set to their ``fiducial'' values.}
        
        Both TDEs belong to the TDE-Bowen spectroscopic class in the \citet{VanVelzen2020} taxonomy, as they display broad Balmer and helium emission, along with complex of lines due to Bowen fluorescence. TDE-Bowen objects present the strongest case for reprocessing, as Bowen fluorescence requires both large densities and a high flux of EUV photons. As noted earlier, none of our models produce pure Balmer emission spectra, so we have not made any comparisons to TDE-H objects, such as AT2018zr. We also do not make any comparisons to TDE-He objects. These transients are probably associated with the disruption of helium-rich stars, so their pure-He spectra are likely to be an abundance effect. {Note, however, that the disruption of a regular main sequence can in principle also result in hydrogen free spectra as a result of ionization effects \citep{Guillochon_2014,Roth2016}.}
        
        Qualitatively, the models do a good job for matching both of the observed spectra in Fig. \ref{fig: tde_model_comparison}. They exhibit the same recombination features, and some with similar line strengths and line profiles, as well as the optical continua which agree quite well. From a more quantitative perspective, there is good agreement between the fiducial model and ASASSN-14li's continuum and emission line spectrum. The only feature(s) missing entirely from the model are Bowen fluorescence feature, which (as noted previously) is produced by a process that is not yet included in \pythonstop. {The agreement is, of course, far from perfect: for example, the Balmer and the \atomictransition{He}{ii} emission lines are too weak, and the continuum is slightly too blue at UV wavelengths.} 
        
        {There is, however, potential for improving the fit of this model by fine-tweaking the model parameters. For example, parameters which modify the density structure of the wind will change the strength of the Balmer lines, since denser winds tend to produce stronger emission as well as brighter optical continua (see \S \ref{sec: formation_of_optical_lines} and Fig. \ref{fig: optical_line_spectra}). Additionally, emission lines can also change by increasing/decreasing the volume of the line emitting region. This could be done by, for example, increasing the covering factor of the wind, i.e. the solid angle subtended by the wind as seen by the central engine, or increasing the outflow outer radius $R_{\rm max}$.  Finally, the continuum SED could also be changed to something more realistic, which better approximates the underlying emission in ASASSN14-li. This would likely cause changes to both the line profiles, as well as the overall shape of the continuum.}
        
        {The match between model and observation is slightly less good for AT2019qiz. Here, the synthetic spectrum is slightly redder than the data. Increasing the mass accretion rate of the disc (and scaling the mass loss of the wind appropriately) would probably fix this and make the spectrum slightly bluer. But the model produces the similar emission features, although the Balmer lines are significantly weaker than the data, with the higher order Balmer features missing entirely.}
        
        Overall, given that we have not fine-tuned our models to precisely match the data, we consider the qualitative agreement with observations extremely promising. More specifically, it confirms that reprocessed disc emission can naturally produce both the continuum and emission lines that are commonly seen in the optical spectra of TDE-Bowen objects.
        
    \subsection{UV spectral lines as geometry and orientation indicators} \label{section: pcygni_discussion}

        \begin{figure*}
            \centering
            \includegraphics[scale=0.55]{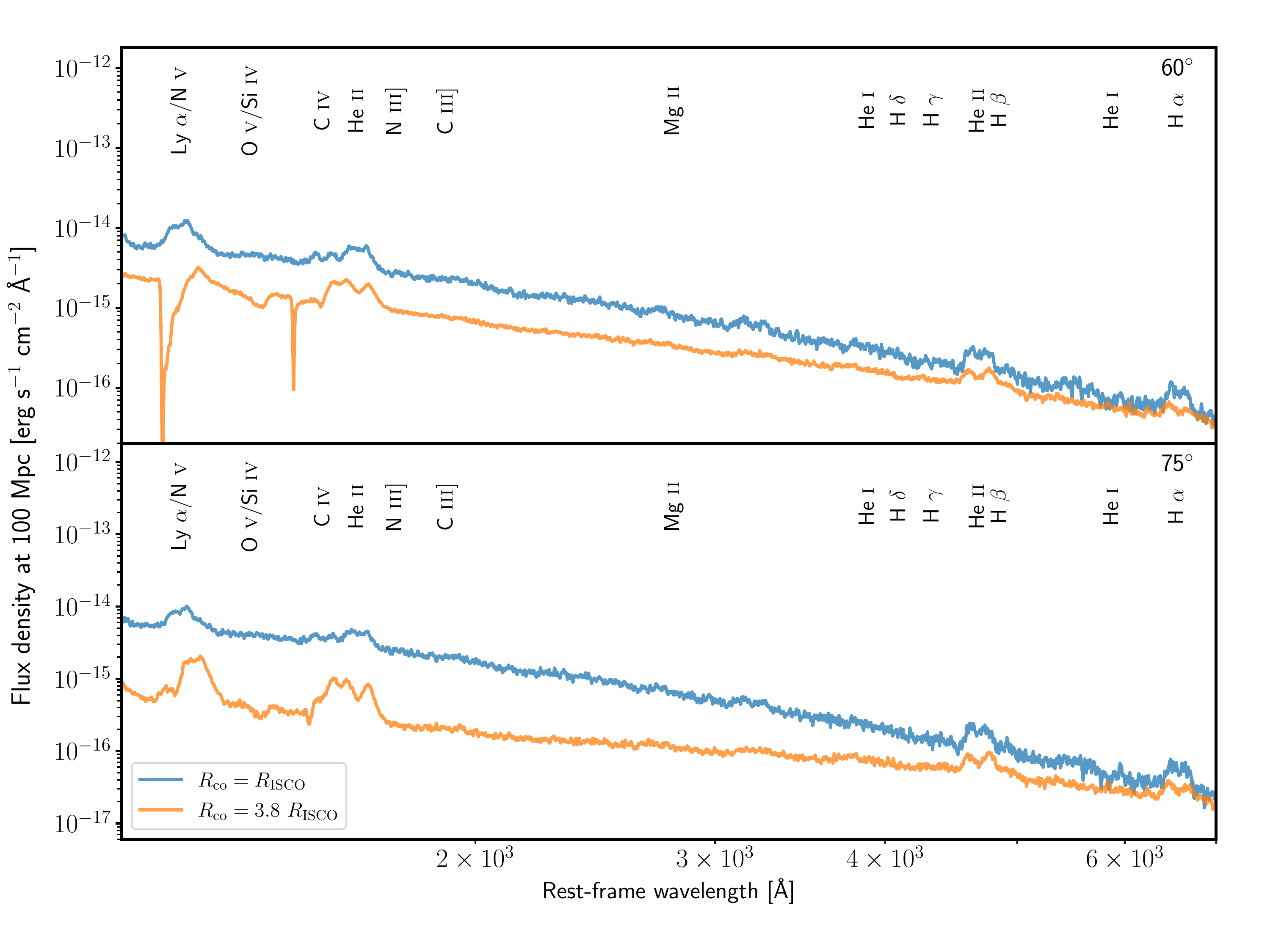}
            \caption{Synthetic spectra of the fiducial model (blue) and a modified fiducial model with a truncated inner disc radius (orange, $R_{\rm co} = 3.8~R_{\rm ISCO}$). The fiducial model in this work (blue) is too highly ionized to produce the UV BAL features reported by \citetalias{Parkinson2020}. By truncating the inner disc radius of the fiducial model, the disc SED is softened and the ionization state of the wind is lowered. The result is that the BEL vs. BAL behaviour reappears for the truncated disc model (orange). Absorption features, such as in C \textsc{iv}, form for sight lines which look into the wind cone (top panel)  and emission features form for sight lines with do not look into the wind cone (bottom panel). \textit{Top panel:} an intermediate sight line looking into the wind cone ($60^{\circ}$). \textit{Bottom panel:} a sight line looking below the wind cone ($75^{\circ}$). }
            \label{fig: pcygni_grid}
        \end{figure*}

        In \citetalias{Parkinson2020}, we showed how the BEL vs. BAL dichotomy found in the UV spectra of TDEs \citep[see the top panel in Fig. \ref{fig: dual_p_tde_uv_opt_obs}, and, e.g.,][]{Cenko_2016, Blagorodnova_2019, Hung2019, hung_discovery_2020} could be explained via line formation in an accretion disc wind. In this picture, BALs are preferentially produced for sight lines which look "into" the wind cone. For other orientations, BELs are observed. The fraction of BAL vs. BEL TDEs can then provide an estimate of the covering fraction of outflows in TDEs. Just as importantly, UV lines could be used to infer our orientation with respect to the TDE disc plane.

        The models presented in this work all produce the UV transitions commonly see in the spectra of TDEs. However, they do not have the same inclination dependence as the synthetic UV spectra in \citetalias{Parkinson2020}. Specifically, none of the in-wind sight lines give rise to BALs in our models. Instead, BELs are produced for all inclinations. 
        
        The absence of BALs in the UV spectra of our models is due to their higher ionization state (relative to the models explored by \citetalias{Parkinson2020}). This is the result of the lower black hole masses we consider, which give rise to harder SEDs. For example, in the fiducial model of \citetalias{Parkinson2020}, the adopted black hole mass is $\text{M}_{\rm BH} = 3 \times 10^{7}~\text{M}_{\odot}$. By contrast, our fiducial model here assumes a black hole mass of $\text{M}_{\rm BH} = 3 \times 10^{6}~\text{M}_{\odot}$. As discussed in Section \ref{sec: model_grid_setup}, this choice, and the range we explore, is based on the data in Fig. \ref{fig: Mbh_parameter_space}. 
        
        Given that BALs are observed in the UV spectra of at least some TDEs, it is worth asking whether disc wind models could still produce these features, even at the lower black hole masses we have considered here. Fundamentally, the answer is clearly yes. For example, the ionization state of a wind model can be lowered by stronger clumping (i.e. a lower volume filling factor and denser clumps), by self-shielding \citep[see, e.g.][]{Murray1995, Proga2000, Proga2004}, or by a softer underlying SED. 
        
        How significant a change is required in order for our fiducial model to produce BALs? In order to address this question, we have carried out tests with both lower filling factors and modified SEDs. These revealed that a reduction in the filling factor by a factor of 10, to $f_V = 0.01$, is not always sufficient enough to produce BALs. By contrast, only a modest change to the disc SED is required before absorption reliably reappears in the UV spectrum. 
    
    	{The modified SED test is illustrated in Fig. \ref{fig: pcygni_grid}. For this test, the inner disc radius was simply truncated to $3.8~R_{\rm ISCO}$, which decreases the maximum temperature and softens the SED. This models illustrates that these models can recover the BEL vs. BAL behaviour, even if the absorption is significantly narrower here than found by \citetalias{Parkinson2020}. More specifically, absorption is seen for inclinations looking into the wind cone ($60^{\circ}$), whereas only BELs are observed for sight lines which do not look into the wind cone ($75^{\circ}$ in Fig. \ref{fig: pcygni_grid}).}
    
        Our blackbody $\alpha$-disc SED is clearly a highly idealized description of the spectrum produced by the central engine (see also Section \ref{sec: limitations} below). Since BAL formation in our models depends sensitively on the shape of this spectrum, the presence or absence of these features cannot be used as a "pure" orientation indicator. However, the fundamental point made by \citetalias{Parkinson2020} remains: BALs can only be observed for specific sight lines, typically those looking into the wind cone. 
        
    \subsection{Limitations} \label{sec: limitations}
        
        As is the case for any numerical simulation, our modelling involves a number of limiting assumptions and simplifications. First, whilst \python now takes into account special relativistic effects, we still neglect general relativistic effects, i.e. whilst \python simulates the full random walk of photons, they still travel along straight lines between interaction points. We also do not include any concept of an event horizon within the computational domain. In \citetalias{Parkinson2020}, we partially tested the importance of general relativistic effects by imposing an absorbing spherical boundary at $R_{\text{ISCO}}$. The ionization state and the emergent spectrum were changed only marginally in this test. We have repeated this test for our fiducial model here, with the same results. 
        
        Second, all \python simulations are time-independent. We have made the implicit assumption that the outflow is in a steady state, which, especially at early times in the evolution of TDEs, may not be true. More specifically, the steady state assumption is valid so long as photon travel times through the flow are short relative to the time-scale over which the luminosity, the SED or the outflow itself can evolve significantly. 
        
        In some TDEs, i.e. ASASSN-14ae, AT2018zr and AT2019qiz, the optical spectra have been observed to evolve significantly over time-scales as short as 10-20 days at early times. In our modelled outflows, photons typically take up to 1-3 days to escape from the simulation grid. Although, some photons may stay within the wind cone for almost 231 days. However, since we are not modelling the initial super-Eddington phase, our steady state assumption is likely to be quite reasonable. In line with this, simulations have shown that TDEs are able to reach a quasi-steady state after their initial super-Eddington phase \citep[e.g.][]{Dai2018}.
        
        Third, reiterating another limitation mentioned previously, we have assumed that the accretion disc is in a steady state, and is geometrically thin and optically thick. Specifically, we model the accretion disc SED by assuming a steady state $\alpha$-disc, with constant $\dot{\rm M}_{\rm acc}$, that locally emits as a blackbody \citep{Shakura1973}. Strictly speaking, our use of the standard $\alpha$-disc temperature profile -- which assumes $\dot{\rm M}_{\rm acc}$ as a function of radius -- is not self-consistent, since our discs suffer significant mass loss from the surface. In principle, this can be modelled by adjusting the effective temperature distribution \citep[see, e.g.][]{knigge_effective_1999}. Such models have, in fact, already been considered for TDEs. Specifically, \citet{miller_flows_2015} outlined such an idea, constructed from work by \citet{laor_line-driven_2014} for AGN. In this scheme, \citet{miller_flows_2015} show that mass losses from the disc, via a disc wind, reduces the accretion rate at the inner edge of the disc, decreasing the temperature and regulating the thermal emission. In practice, the net effect is likely to be similar to the inner disc truncation we explored in Section \ref{section: pcygni_discussion}. 
        
        Another limitation of our disc model is that the inner regions of TDE discs are dominated by radiation pressure and probably vertically extended. The most important effect of the thin-disc approximation in this context is on the angular distribution of the radiation field. Due to foreshortening and limb darkening, geometrically thin and optically thick discs generate highly anisotropic radiation fields. This degree of anisotropy is likely an overestimate for the vertically extended inner discs in TDEs. However, the structure, evolution and even the stability of such radiation dominated discs is the subject of on-going intense research \citep[e.g.][]{Hirose2009, Jiang_2013, blaes14}. Our choice to use a simple disc model is, then, one borne out of practically. Given that the emission from the disc is reprocessed by the outflow, which isotropizes the radiation field, we do not expect our qualitative results to change significantly.
        
        Fourth and finally, we only model hydrogen and helium self-consistently with full, multi-level model atoms. By contrast, bound-bound transitions in metals are treated using a 2-level atom approximation. Whilst this is perfectly reasonable for resonance lines, it is not reasonable for transitions involving excited and/or meta-stable states. With the exception of C \textsc{iii}] $\lambda$1909 (which we treat in an approximate manner via a modified 2-level atom approximation), our simulations do not include realistic treatments for semi-forbidden transitions such as N \textsc{iii}] $\lambda$1750. Our simplified treatment of metals also currently prevents us from modelling the fluorescence process that gives rise to the Bowen blend between 4630~\AA~and 4660~\AA, commonly seen in observations.
        
\section{Conclusions} \label{sec: conclusion}

    We have shown that the reprocessing of disc radiation by an accretion disc wind can naturally produce the optical line and continuum spectra seen in TDEs. In order to achieve this, we have run a grid of Monte Carlo radiative transfer and ionization simulations to produce synthetic UV and optical spectra of wind- and disc-hosting TDEs. 
    
    The disc winds we model are rotating, biconical and clumpy, and they are illuminated by a geometrically thin and optically thick accretion disc. Our model grid covers a realistic range of wind kinematics, black hole masses and accretion states, inspired by observations. Using this grid, we explore how the mass of the black hole, the accretion rate and the wind properties affect the broadband SED and optical line spectra. Our main results are as follows:
    
    \begin{enumerate}
        \item The hydrogen and helium emission lines commonly seen in the optical line spectra of TDEs can be produced by an optically thick accretion disc wind that reprocesses the SED of the underlying disc. Reprocessing in such an outflow can also naturally produce the UV line spectra seen in TDEs. 
        
        \item Our models produce excellent matches to the optical spectra of TDEs of the TDE-Bowen class, reproducing the correct continuum shapes and emission line properties. However, our models do not produce the pure Balmer spectra seen in the TDE-H spectroscopic class.
        
        \item All of our outflow models are optically thick. There is significant opacity in all of the modelled outflows at the \atomictransition{He}{ii} ($\approx 54$ eV) photo-ionization edge, with typical integrated optical depths of $\tau \sim 10^{7}$ at high and intermediate inclinations. At low inclinations, the optical depth of this edge is still large, but can be of similar magnitude to the electron scattering optical depth. Our models also present significant opacity at the \atomictransition{He}{i} (24.6 eV) and, in some models, the hydrogen Lyman edges. The electron scattering optical depth is similar in all of our models, {with $\tau \simeq 1~-~55$ depending on inclination.}
        
        \item The optical colour temperature of the reprocessed SED is much redder than that of the input accretion disc SED. Changes to the black hole mass, accretion rate and the kinematics of the wind affect the degree of reprocessing and the properties of the line spectra. In general, slower and denser winds result in a greater amount of reprocessing, as well as stronger Balmer and helium recombination features.
        
        \item The main reprocessing mechanism is photo-ionization of \atomictransition{He}{i} and \atomictransition{He}{ii} near the base of the wind. The absorbed luminosity is then re-radiated at longer wavelengths via recombination, or as free-free or line emission. {Based on our simulations, we expect \atomictransition{He}{ii} bound-free interactions to be a critical source of opacity for reprocessing the continuum emission in TDEs. Free-free absorption can also play an important role, but affects the spectra to a lesser degree.}
        
        \item The optical emission lines are often double peaked, since the kinematics of the line-forming regions are dominated by rotation. Some lines exhibit asymmetric red wings due to adiabatic reprocessing. Electron scattering also dramatically affects the line profiles, both by increasing the width of the lines and, for low inclinations, by smoothing out the double peaked shape. 
        
        \item Our models produce synthetic UV spectra with the same set of atomic transitions commonly seen in TDEs. However, we find that the winds are often too highly ionized to produce UV absorption features. Truncating the inner radius of the accretion disc at $\simeq 4~R_{\text{ISCO}}$ softens the SED, lowers the ionization state of the wind and recovers UV broad absorption lines for sight lines that look into the wind cone.
    \end{enumerate}
    
\section*{Data availability}

    The data underlying this article will be shared on reasonable request to the corresponding author, or can be accessed online at \hyperlink{https://github.com/saultyevil/tde_optical_reprocessing}{github.com/saultyevil/tde\_optical\_reprocessing}.
    
\section*{Acknowledgements}

    {We thank the anonymous reviewer for their careful reading of the manuscript, and their helpful and insightful feedback which significantly improved the quality and usefulness of this work.} Figures were prepared using \texttt{matplotlib} \citep{Hunter2007}. The authors acknowledge the use of the IRIDIS High Performance Computing Facility, and associated support services at the University of Southampton. The authors acknowledge the use of the \texttt{GNU Science Library} \citep{gsl}. EJP would like to acknowledge financial support from the EPSRC Centre for Doctoral Training in Next Generation Computational Modelling grant EP/L015382/1. JHM acknowledges a Herchel Smith Research Fellowship at Cambridge. NSH and CK acknowledge support from the Science and Technology Facilities Council grant ST/V001000/1. KSL acknowledges partial support for this project by NASA through grant number HST-GO-15984, 16058 and 16066 from the Space Telescope Science Institute, which is operated by AURA, Inc., under NASA contract NAS 5-26555.



\bibliographystyle{mnras}
\bibliography{myreferences}





\bsp
\label{lastpage}
\end{document}